\documentclass{article}

\usepackage[english]{babel}

\usepackage[letterpaper,top=2cm,bottom=2cm,left=2cm,right=2cm,marginparwidth=1.75cm]{geometry}

\usepackage{mathrsfs}
\usepackage{amsmath}
\usepackage{amssymb}
\usepackage{epsfig}
\usepackage{graphicx}
\usepackage{subfigure}
\usepackage{multirow}
\usepackage{epstopdf}
\usepackage{authblk}
\usepackage{cite}
\usepackage{float}

\def\be{\begin{eqnarray}}
\def\en{\end{eqnarray}}
\def\non{\nonumber\\}

\title{Branching ratios and CP asymmetries of the quasi-two-body decays $B_c \rightarrow \ K^{*}_0(1430,1950) D_{(s)} \rightarrow K \pi D_{(s)} $ in the PQCD approach}
\author{Zhi-Qing Zhang$^1$, Zi-Yu Zhang$^1$, Ming-Xuan Xie$^1$, Ming-Yang Li$^1$, Hong-Xia Guo\footnote{Corresponding author: guohongxia@zzu.edu.cn} $^2$  } 
\affil{ \it \small $^1$  Institute of Theoretical Physics, School of Sciences, Henan University
of Technology,\\ \it Zhengzhou, Henan 450001, China\\
\it \small $^2$  School of Mathematics and Statistics, Zhengzhou University, \\ \it Zhengzhou, Henan 450001, China}
\date{\today}

\begin{document}
\maketitle
\begin{abstract}
In this paper, we investigate the quasi-two-body decays $B_c \to K_0^{*}(1430,1950) D_{(s)} \to K \pi D_{(s)}$ within the perterbative QCD (PQCD) framework. The S-wave two-meson distribution amplitudes (DAs) are introduced to describe the final state interactions of the $K\pi$ pair, which involve the time-like form factors and the Gegenbauer polynomials. In the calculations, we adopt two kinds of parameterization schemes to describe the time-like form factors: One is the relativistic Breit-Wigner (RBW) formula, which is usually more siutable for the narrow resonances, and the other is the LASS line shape proposed by the LASS Collaboration, which includes both the resonant and nonresonant components. We find that the branching ratios and the direct CP violations for the decays $B_c \to K_0^{*}(1430) D_{(s)}$ obtained from those of  the quasi-two-body decays $B_c \to K_0^{*}(1430) D_{(s)} \to K \pi D_{(s)}$ under the narrow width approximation (NWA) can be consistent well with the previous PQCD results calculated in the two-body framework by assuming that $K^*_0(1430)$ is the lowest lying $\bar q s$ state, which
is the so-called scenario II (SII). We conclude that the LASS parameterization is more siutable to describe the $K_0^{*}(1430)$ than the RBW formula, and the nonresonant components play an important role in the branching ratios of the decays $B_c \to K_0^{*}(1430) D_{(s)} \to K \pi D_{(s)}$. In view of the large difference between the decay width measurements for the $K_0^{*}(1950)$ given by BaBar and LASS collaborations, we calculate the branching ratios and the CP violations for the quasi-two-body decays $B_c \to K_0^{*}(1950) D_{(s)} \to K \pi D_{(s)}$ by using two values, $\Gamma_{K^*_0(1950)}=0.100\pm0.04$ GeV  and $\Gamma_{K^*_0(1950)}=0.201\pm0.034$ GeV, besides the two kinds of parameterizations for the resonance $K^*_0(1950)$. We find that the branching ratios and the direct CP violations for the decays $B_c \to K_0^{*}(1950) D_{(s)} \to K \pi D_{(s)}$ have not as large difference between the two  parameterizations as the case of decays involving the $K^*_0(1430)$, especially for the results with $\Gamma_{K^*_0(1950)}=0.201\pm0.034$ GeV. The effect of the nonresonant component in the $K^*_0(1950)$ may be not so serious as that in the $K^*_0(1430)$.  The quasi-two-body decays $B^+_c \to K^{*+}_0(1430) D^{0} \to K^0 \pi^+ D^{0}$
and $B^+_c \to K^{*0}_0(1430) D^{+} \to K^+ \pi^- D^{+}$ have large branching ratios, which can reach to the order of  $10^{-4}$ and are most likely to be observed in the future LHCb experiments. Furthermore, the branching ratios of the quasi-two-body decays 
$B_c \to K_0^{*}(1950) D_{(s)} \to K \pi D_{(s)}$ are about one order smaller thant those of the corresponding decays $B_c \to K_0^{*}(1430) D_{(s)} \to K \pi D_{(s)}$. 
\end{abstract}
{\centering\section{INTRODUCTION}\label{intro}}

Recently, a lot of charmed three-body b-flavored heavy meson ($B,B_s, B_c$) decays have attract many attentions on the experimental side. For example, the decays $B^0\to \bar D^{(*)0} K^+\pi^-$ \cite{lhcb1,lhcb2},
$B_s\to \bar D^{(*)0} K^-\pi^+$ \cite{lhcb2,lhcb3}, $B_s\to \bar D^0 K^+K^-$ \cite{lhcb4} and $B^0\to D^+\pi^-\pi^-$ \cite{lhcb44} have been measured in detail at LHCb. Many of these channels have large branching ratios, which are in the range $10^{-5}\sim 10^{-4}$. Besides being used to extract the unitary triangles in the standard model (SM), they provide a platform for probing the rich resonance structures and studying the localized CP violations. Many resonances have been found and studied in the such decays, for example, six $K\pi$ resonances were searched in the decay $B_s^0\to\bar D^0 K^-\pi^+$ \cite{lhcb2}. It is helpful to understand the properties and inner structures of these resonances. As we know, the resonant contributions are commonly described by the relativistic Breit-Wigner (RBW) line shape, while such RBW parameterization is invalid for the broad-width resonances, such as $K^*_0(1430)$, which interferes strongly with a slowly varying nonresonant background \cite{meadows}.
Then the so-called LASS lineshape \cite{lass} was suggested by the LASS collaboration to describe this resonance. In practice, different collaborations adopted different formulas to decribe the $K^*_0(1430)$. For example, Belle employed the RBW model 
and an exponential parameterization to describe the resonant and nonresonant contributions in the $K^*_0(1430)$, respectively \cite{bellek}, while BaBar used the LASS parameterization to describe the $K^*_0(1430)$ resonance \cite{babark}. On the theoretical side, different parameterizations for the $K^*_0(1430)$ were also adopted by different authors. For example, the RBW formula was adopted  to describe the  $K^*_0(1430)$ in the decays $B\to K^*_0(1430)h \to K \pi h$ with $h$ being $\pi, K$ \cite{wfwang}, while the LASS parameterization was used in the decays $B\to K^*_0(1430)D\to K\pi D$ \cite{wsfang}.    Another scalar resonance $K^*_0(1950)$ with a little narrower width $\Gamma_{K^*_0(1950)}=201\pm34$ MeV was observed in the LASS experiment for the reaction $K^-p\to K^-\pi^+n$ \cite{lass}, where it was considered as a radial excitation of the $0^+$ member with $L=1$ triplet. Recently, its decay width was updated to $\Gamma_{K^*_0(1950)}=80\pm38$ MeV by BaBar \cite{babar1}. The underlying structures of the scalar mseons are long-standing puzzle and the classification of them is one of most interesting topics in hadron physics. Usually, there are two typical
schemes for their classification: The nonet mesons below 1 GeV are viewed as the lowest lying $q\bar q$ states, while the nonet ones near 1.5 GeV including $K^*_0(1430)$ are suggested as the first excited states. It is called scenario I (SI). While in scenario II (SII), the nonet mesons near 1.5 GeV are treated as $q\bar q$ ground states, and the nonet mesons below 1 GeV are exotic states, such as four-quark bound states. 

The $B_c$ meson is the lowest-lying bound state of a bottom antiquark and a charm quark with $J^P=0^{-}$. Because that both of its constituent quarks are heavy and can decay individually, a very rich $B_c$-decay channels with sizable branching ratios are expected, which will provide a unique platform for studying the weak decay mechanism of heavy flavor mesons. Since more than $10^{10}$ $B_c$ events per year can be collected at LHC with the luminosity of $\mathcal{L}=10^{34} \mathrm{~cm}^{-2} \mathrm{~s}^{-1}$ \cite{gouz} at present, we will enter the precision era in $B_c$ physics in the near future. Measurements of some $B_c$ meson three-body decays, such as $B_c\to K^+K^-\pi^+$ \cite{lhcb5}, $B_c\to J/\Psi D^{(*)0}K^+, J/\Psi D^{(*)+}K^{*0}$ \cite{lhcb6} have been performed.
 Furthermore, the two-body decays $B_c\to K^*_0(1430)D_{(s)}$ have been studied in previous work \cite{Zou:2017yxc}, where the branching ratios are predicted as $10^{-5}\sim 10^{-4}$. So large branching ratios can be detected by the future LHCb experiments. In view of upper listed reasons, we would like to study the scalar resonances $K^*_0(1430)$ and $K^*_0(1950)$ through the quasi-two-body decays $B_c \rightarrow \ K^{*}_0(1430,1950) D_{(s)} \rightarrow K \pi D_{(s)}$ within the PQCD framework.
 In order to study $B_{(c)}$ meson three-body decays, many approaches based on symmetry principles and factorization theorems have been proposed. Symmetry principles incude the U-spin \cite{Bhat,Gronau,xu}, isospin and flavor $SU(3)$ symmetry \cite{Gronau2,Engelhard,Imbeault,He}, and the factorization-assisted topological-diagram amplitude (FAT) approach \cite{Zhou}, etc. Factorization theorems include the QCD-improved factorization approach \cite{Krankl,Cheng,Li,Cheng2007,Klein} and the PQCD approach \cite{zhao, zhang, zhang2, Liy, maaj}. It has been proposed that the factorization theorem of the quasi-two-body $B_{(c)}$ decays is approximately valid when the two particles move collinearly and the bachelor particle recoils back in the final states. This case corresponds to the edges of the Dalitz plot, which provides us a great opportunity to probe the properties of various resonances. According to this quasi-two-body-decay mechanism, the two-meson distribution amplitudes (DAs) are introduced into the PQCD approach, where the strong dynamics between the two final hadrons in the resonant regions are included.

This paper is organized as follows. The framework of the PQCD approach for the quasi-two-body $B_{c}$ decays is reviewed in Section II,
where the kinematic variables for each meson are defined and the S-wave $K\pi$ pair distribution amplitudes up to twist-3 are parametrized.  Then, the analytical formulas of each Feynman diagram and the total amplitudes for these decays are listed.
In Section III, the numerical results and discussions are presented. The final section is devoted to our conclusions. Some
details related functions are collected in the Appendix.

{\centering\section{THE FRAMEWORK }}

In the framework of the PQCD approach for the quasi-two-body decays, the factorization formulas for the $B_c \rightarrow \ K^{*}_0(1430,1950) D_{(s)} \rightarrow K \pi D_{(s)}$
decay amplitudes can be written as \cite{Chen:2002th,Chen:2004az}
\be
\mathcal{A}=\Phi_{B_c} \otimes H \otimes \Phi^{\text{S-wave}}_{K\pi} \otimes \Phi_{D_{(s)}},
\en
where $\Phi_{B_c}(\Phi_{D_{(s)}})$ denotes the DAs of the initial (final bachelor) meson, $\Phi^{\text{S-wave}}_{K\pi}$ is the S-wave $K\pi$ pair DAs, and
$\otimes$ denotes the convolution integrations over the parton momenta. Similar to the two-body decay case, the evolution of the hard kernel H for the b quark decay is calculable perturbatively and starts with the diagrams of single hard gluon exchange. The nonperturbative dynamics are absorbed into those DAs $\Phi_{B_c}, \Phi_{D_{(s)}}$ and $\Phi^{\text{S-wave}}_{K\pi}$.

In the rest frame of the $B_c$ meson, we define the $B_c$ meson momentum $p_{B_c}$, the $K (\pi)$ meson momentum $p_1 (p_2)$, the scalar meson $K^*_0$\footnote{From now on we use $K^*_0$ to reprent $K^*_0(1430)$ or $K^*_0(1950)$ for simplicity. }  momentum $p=p_1+p_2$, and the bachelor meson $D_{(s)}$ momentum $p_3$ in light-cone coordinates as
\be
   p_{B_c}&=&\frac{m_{B_c}}{\sqrt{2}}\left(1,1, \mathbf{0}_{\mathrm{T}}\right), \quad p=\frac{m_{B_c}}{\sqrt{2}}\left(\zeta, 1-r^2, \mathbf{0}_\mathrm{T}\right), \quad p_3=\frac{m_{B_c}}{\sqrt{2}}\left(1-\zeta, r^2, \mathbf{0}_\mathrm{T}\right),
\en
where the mass ratio $r=m_{D_{(s)}}/m_{B_c}$ and $\zeta=s/m^2_{B_c}$ with the invariant mass square $s=p^2=m^2_{K\pi}$ for the $K\pi$ pair. The momenta of the light quarks in the initial meson $B_c$, the scalar meson $K^*_0$ and the bachelor meson $D_{(s)}$ are defined as $k_1, k$ and $k_3$, respectively
\begin{equation}
    k_1= \frac{m_{B_c}}{\sqrt{2}}\left(x_1,0, \mathbf{k}_{1\mathrm{T}}\right), \quad k=\frac{m_{B_c}}{\sqrt{2}} \left(0, (1-r^2)z, \mathbf{k}_{\mathrm{T}}\right), \quad k_3=\frac{m_{B_c}}{\sqrt{2}}\left((1-\zeta)x_3, 0, \mathbf{k}_{3\mathrm{T}}\right),
\end{equation}
where $x_1$, $z$ and $x_3$ are the corresponding momentum fractions.

{\centering\subsection{WAVE FUNCTIONS }}

In the course of the PQCD calculations, the necessary inputs contain the DAs, which are constructed via the nonlocal matrix elements. Since the $B_c$ meson is composed of two heavy quarks, one can take the nonrelativistic approximation form, that is the zero-point wave function, for the $B_c$ meson light-cone distribution amplitudes (LCDAs)
\begin{equation}
\Phi_{B_c}(x)=\frac{i f_{B_c}}{4 N_c}\left[\left(P\hspace{-2.1truemm}/+m_{B_c}\right) \gamma_5 \delta\left(x-r_c\right)\right] \exp \left(-\frac{\omega_{b}^2 b^2}{2}\right),
\end{equation}
where $N_c=3$ is the color factor, the mass ratio $r_c=m_c/m_{B_c}$ and the shape parameter  $\omega_b = 0.6 \pm 0.05$ GeV. Here, only the Lorentz structure offering the dominant contribution is considered, while the contribution from the other one is numerically small and can be neglected. The exponent term describes the intrinsic $k_T$-dependence with $b$ being the conjuate space coordinate.  

For D meson, the two-parton LCDAs in the heavy quark limit can be written as \cite{kuri,rhli}
\begin{equation}
    \left\langle D\left(p\right)\left|q_\alpha(z) \bar{c}_\beta(0)\right| 0\right\rangle=\frac{i}{\sqrt{2N_c}} \int_0^1 d x e^{i x p \cdot z}\left[\gamma_5\left(p\hspace{-1.5truemm}/ +m_D\right) \phi_D(x, b)\right]_{\alpha \beta},
\end{equation}
with the distribution amplitude $\phi_D(x, b)$,
\begin{equation}
    \phi_D(x, b)=\frac{1}{2 \sqrt{2N_c}} f_D 6 x(1-x)\left[1+C_D(1-2 x)\right] \exp \left[\frac{-\omega^2 b^2}{2}\right],
\end{equation}
where $x$ is the momentum fraction of the light quark in $D$ meson, $C_D=0.5 \pm 0.1, \omega=0.1$ GeV and $f_D=211.9$ MeV. It is similar for the LCDAs of $D_s$ meson but with different parameters $C_{D_s}=0.4 \pm 0.1, \omega=0.2$ GeV and $f_{D_s}=249$ MeV, caused by a little SU(3) breaking effect \cite{maaj}.

The $S$-wave $K \pi$ system LCDAs are written as
\begin{equation}
	\Phi_{K \pi}(z, s)=\frac{1}{\sqrt{2 N_c}}\left[p \hspace{-1.5truemm}/\phi_0(z, s)+\sqrt{s} \phi_s(z, s)+\sqrt{s}(\not v\not n-1) \phi_t(z, s)\right],
\end{equation}
where the dimensionless vectors $v=\left(0,1,0_{\mathrm{T}}\right)$ and $n=\left(1,0,0_{\mathrm{T}}\right)$. The twist-2 LCDA is defined as
\begin{equation}
	\phi_0(z, s)=\frac{F_{K \pi}(s)}{2 \sqrt{2 N_c}}\left\{6 z(1-z)\left[a_0(\mu)+\sum_{m=1}^{\infty} a_m(\mu) C_m^{3 / 2}(2 z-1)\right]\right\},
\end{equation}
where $C_m^{3 / 2}$ are the Gegenbauer polynomials,  $a_0=\left(m_s(\mu)-m_q(\mu)\right) / \sqrt{s}$ for $K_0^{*-}, \bar{K}_0^{* 0}$  and $a_0=$ $\left(m_q(\mu)-m_s(\mu)\right) / \sqrt{s}$ for $K_0^{*+}, K_0^{* 0}$ \cite{Cheng:2013fba}. The scale-dependent Gegenbauer moments  $a_1=-0.57 \pm 0.13$ and $a_3=-0.42 \pm 0.22$ at the scale $\mu=$ $1 \mathrm{GeV}$ for the resonance $K_0^*(1430)$, and the contributions from the even terms have been neglected \cite{Cheng:2005nb}. Since the Gegenbauer moments for the resonance $K_0^*(1950)$ are not available at present, we will use the same Gegenbauer moments $a_1$ and $a_3$ with those for $K_0^*(1430)$ in the numerical calculations. As for the twist-3 LCDAs, we adopt the asymptotic forms as follows

\begin{equation}
	\phi_s(z, s)=\frac{F_{K \pi}(s)}{2 \sqrt{2 N_c}}, \quad \phi_t(z, s)=\frac{F_{K \pi}(s)}{2 \sqrt{2 N_c}}(1-2 z).
\end{equation}
Here the time-like form factor $F_{K \pi}(s)$ is related to scalar form factor $F_0^{K \pi}(s)$ by the formula
\be
F_{K \pi}(s)=\frac{m^2_K-m^2_\pi}{m_{K_0^*}(m_s-m_q)} F_0^{K \pi}(s),
\en
where $q$ represents the light quark $u$ or $d$, the scalar form factor $F_0^{K \pi}(s)$ is defined as \cite{Jamin:2001zq,Boito:2009qd,Meissner:2013hya}
\begin{equation}
	\langle K \pi|\bar{q} s| 0\rangle= C_X \frac{m^2_K-m^2_\pi}{(m_s-m_q)}F_0^{K \pi}(s),
\end{equation}
with the isospin factor $C_X=1 (1/\sqrt{2})$ for the $K^{+} \pi^{-}, K^0 \pi^{+} (K^{+} \pi^0, K^0 \pi^0)$ pairs. For the case of the $K^{+} \pi^{-}$ pair originated from the resonant state $K_0^{*0}(1430)$, we can insert a complete of  $K_0^{*0}(1430)$ intermediate state into above matrix elements \cite{Boito:2009qd}
\begin{equation}
	\left\langle K^{+} \pi^{-}|\bar{d} s| 0\right\rangle_{K^*_0} \approx\left\langle K^{+} \pi^{-} \mid K_0^{* 0}\right\rangle \frac{1}{\mathcal{D}_{K_0^*}}\left\langle K_0^{* 0}|\bar{d} s| 0\right\rangle=\frac{g_{K_0^* K \pi}}{m_{K_0^*}^2-s-i m_{K_0^*} \Gamma(s)} m_{K^*_0}\bar f_{K^*_0},
\end{equation}
where the definations of the scalar decay constant $\bar{f}_{K_0^*}$ and the coupling constant $g_{K_0^* K \pi}$, i.e.,  $\left\langle K_0^{* 0}|\bar{d} s| 0\right\rangle=m_{K_0^*} \bar{f}_{K_0^*}$ and $g_{K_0^* K \pi}=\left\langle K^{+} \pi^{-} \mid K_0^{* 0}\right\rangle$, have been used. The scale-dependent scalar decay consant $\bar{f}_{K_0^*}$ is related with the vecter decay constant
$f_{K_0^*}$ by the equations of motion $\bar{f}_{K_0^*}=\frac{m_{K^*_0}f_{K_0^*}}{m_s(\mu)-m_d(\mu)}$.  
We will employ $f_{K_0^*(1430)} m_{K_0^*(1430)}^2=0.0842 \pm 0.0045 \mathrm{GeV}^3$\cite{Maltman:1999jn} and $f_{K_0^*(1950)} m_{K_0^*(1950)}^2=0.0414 \mathrm{GeV}^3$ \cite{Shakin:2001sz} in the following calculations. The mass-dependent decay width $\Gamma(s)$ in the denominator is related with the total decay width $\Gamma_0$ of $K^*_0$ by the formula $\Gamma(s)=\Gamma_0 \frac{q}{q_0} \frac{m_{K_0^*}}{\sqrt{s}}$ with $q$ being the magnitude of the momentum for the daughter meson $K$ or $\pi$
\begin{equation}
	q=\frac{1}{2} \sqrt{\left[s-\left(m_K+m_\pi\right)^2\right]\left[s-\left(m_K-m_\pi\right)^2\right] / s} .
\end{equation}
The $q_0$ in $\Gamma(s)$ is the value for $q$ at $s=m_{K_0^*}^2$. The coupling constant $g_{K_0^* K \pi}$ is determined from 
the partial width $\Gamma_{K_0^* \rightarrow K \pi}$ for the decay $K_0^* \rightarrow K \pi$ through the relation \cite{cheng1}
\begin{equation}
	g_{K_0^* K \pi}=\sqrt{\frac{8 \pi m_{K_0^*}^2 \Gamma_{K_0^* \rightarrow K \pi}}{q_0}}.
\end{equation}

It is noticed that the LASS parametrization is often used to describe the $S$-wave $K \pi$ system, especially for $K^*_0(1430)$ in experiments \cite{BaBar:2005qms}, 
\begin{equation}
R(s)=\frac{\sqrt{s}}{q \cot \delta_B-i q}+e^{2 i \delta_B} \frac{m_{K_0^*} \Gamma_0 \frac{m_{K_0^*}}{q_0}}{m_{K_0^*}^2-s-i m_{K_0^*} \Gamma_0 \frac{q}{\sqrt{s}} \frac{m_{K_0^*}}{q_0}},
\label{lass}
\end{equation}
where $\cot \delta_B=\frac{1}{a q}+\frac{1}{2} r q$ with the scattering length $a=2.07 \pm 0.10 \mathrm{GeV}^{-1}$ and the effective range $r=3.32 \pm 0.34 \mathrm{GeV}^{-1}$ \cite{BaBar:2005qms}. The first (second) term refers to the nonresonant (resonant) contribution. If using the LASS parametrization to replace the RBW formula in the sclalar form factor $F^{K\pi}(s)$, one can get the time-like form factor $F_{K\pi}(s)$ 
\begin{equation}
F_{K \pi}(s)=\frac{q_0}{m_{K_0^*}^2 \Gamma_0} g_{K_0^* K \pi} \bar{f}_{K_0^*} R(s).
\end{equation}
For comparison, these two parametrizations are used in the numerical calcualtions.
  
{\centering\subsection{ Analytic formulae } \label{formu}}

For the quasi-two-body decays $B_c \rightarrow \ K^{*}_0 D_{(s)} \rightarrow K \pi D_{(s)}$, the effective Hamiltonian relevant to the $b \to s(d) $ transition is given by \cite{heff}
\begin{equation}
\begin{aligned}
H_{e f f}= & \frac{G_F}{\sqrt{2}}\left\{\sum_{q=u, c} V_{q b} V_{q s(d)}^*\left[C_1(\mu) O_1^{(q)}(\mu)+C_2(\mu) O_2^{(q)}(\mu)\right]\right. \\
& \left.-\sum_{i=3 \sim 10} V_{t b} V_{t s(d)}^* C_i(\mu) O_i(\mu)\right\}+H . c .,
\end{aligned}
\end{equation}
where the Fermi coupling constant $G_F \simeq 1.166 \times 10^{-5} \mathrm{GeV}{ }^{-2}$, $V_{q b} V_{q d(s)}^*$ and $V_{t b} V_{t d(s)}^*$ are the products of the Cabibbo-Kobayashi-Maskawa (CKM) matrix elements. According to the scale $\mu$, one can separate the effective Hamiltonian into two distinct parts: the Wilson coefficients $C_i$, and the local four-quark operators $O_i$. The local four-quark operators for the $b\to d$ transition are written as

\begin{figure}[htbp]
\centering
        \includegraphics[width=1\textwidth]{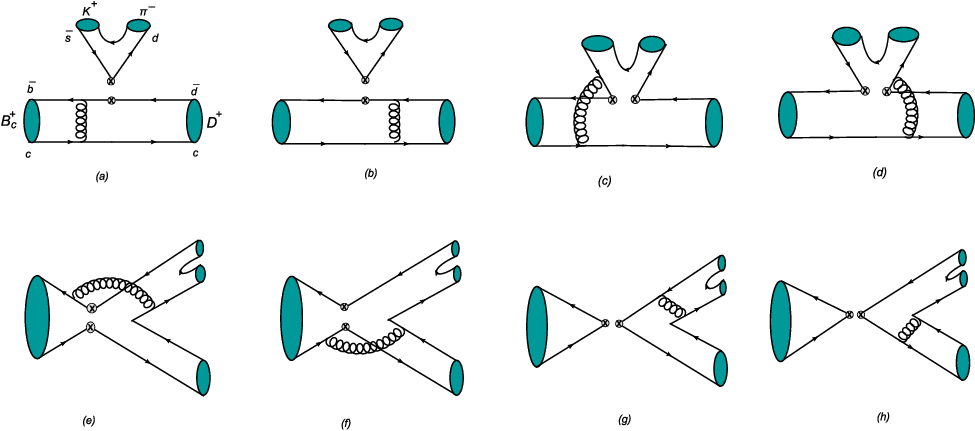}

          \caption{The leading order Feynman diagrams for the decay $B_c^+ \rightarrow K^{*0}_0D^{+} \rightarrow K^{+}\pi^- D^{+}$. }
        \label{feynman}
   \end{figure}
\begin{equation}
\begin{aligned}
    O_1^{(q)} & =\left(\bar{d}_i q_j\right)_{V-A}\left(\bar{q}_j b_i\right)_{V-A}, & O_2^{(q)} &=\left(\bar{d}_i q_i\right)_{V-A}\left(\bar{q}_j b_j\right)_{V-A}, \\
O_3 & =\left(\bar{d}_i b_i\right)_{V-A} \sum_q\left(\bar{q}_j q_j\right)_{V-A}, & O_4 &=\left(\bar{d}_i b_j\right)_{V-A} \sum_q\left(\bar{q}_j q_i\right)_{V-A}, \\
O_5 & =\left(\bar{d}_i b_i\right)_{V-A} \sum_q\left(\bar{q}_j q_j\right)_{V+A}, & O_6 &=\left(\bar{d}_i b_j\right)_{V-A} \sum_q\left(\bar{q}_j q_i\right)_{V+A}, \\
O_7 & =\frac{3}{2}\left(\bar{d}_i b_i\right)_{V-A} \sum_q e_q\left(\bar{q}_j q_j\right)_{V+A}, & O_8 &=\frac{3}{2}\left(\bar{d}_i b_j\right)_{V-A} \sum_q e_q\left(\bar{q}_j q_i\right)_{V+A}, \\
O_9 & =\frac{3}{2}\left(\bar{d}_i b_i\right)_{V-A} \sum_q e_q\left(\bar{q}_j q_j\right)_{V-A}, & O_{10} &=\frac{3}{2}\left(\bar{d}_i b_j\right)_{V-A} \sum_q e_q\left(\bar{q}_j q_i\right)_{V-A},
\label{operator}
\end{aligned}
\end{equation}
where $\mathrm{i, j}$ are the color indices and $V \pm A$ refer to the Lorentz structures $\gamma_\mu\left(1 \pm \gamma_5\right)$. Similarly, the local four-quark operators for the $b\to s$ transtion can be obtained by replacing $d$ with $s$
in Eq. (\ref{operator}).

The typical Feynman diagrams at the leading order for the quasi-two-body
decays $B_c \rightarrow \ K^{*}_0 D_{(s)} \rightarrow K \pi D_{(s)}$ are shown in Fig. \ref{feynman}, where we take the decay
$B_c^+\to K^{*0}_0D^{+} \to K^{+}\pi^{-} D^{+}$ as an example.
We mark LL, LR, SP to denote the contributions from $(V-A)(V-A),(V-A)(V+A)$ and $(S-P)(S+P)$ operators, respectively. The amplitudes from the factorizable emission diagrams Fig. 1(a) and 1(b) are given as
\be
\mathcal{F}_e^{L L}&=& 2 \sqrt{\frac{2}{3}} \pi f_{B_c} F_{K \pi}m_{B_c}^4 C_F \int_0^1  d x_3 \int_0^{\infty} b_1 b_3 d b_1 d b_3 \exp(-\frac{\omega^2_bb^2_1}{2}) \phi_D(x_3,b_3) \non
&&\times\left\{ \left[-\zeta+r\left(4 \zeta-(\zeta+1) r_b-2(\zeta-1) x_3\right)+2 r_b(1-\zeta)- (1-\zeta)^2x_3\right]\right.\non
&&\left.\times\alpha_s (t_a) h\left(\alpha_e, \beta_a, b_1, b_3\right)  \exp \left[-S_{a b}\left(t_a\right)\right]S_t(x_3)\right. \non
&&\left.-[2r\left( \zeta\left(r_c+1\right)+r_c-1\right)-\zeta r_c]\right.\non
&&\left.\times\alpha_s (t_b) h\left(\alpha_e, \beta_b, b_1, b_3\right)  \exp \left[-S_{a b}\left(t_b\right)\right]S_t(r_c)\right\},\\
\mathcal{F}_e^{LL}&=& \mathcal{F}_e^{LR},\\
\mathcal{F}_e^{SP}&=&-4 \sqrt{\frac{2}{3}} \pi f_{B_c} F_{K \pi}m_{B_c}^4 C_F\sqrt{\zeta} \int_0^1  d x_3 \int_0^{\infty} b_1 b_3 d b_1 d b_3 \exp(-\frac{\omega^2_bb^2_1}{2}) \phi_D(x_3,b_3) \non
&&\times\left\{ \left[r(\zeta(1-x_3)-4r_b+x_3+1)-(1-\zeta)(2-r_b)\right]\right.\non
&&\left.\times\alpha_s (t_a) h\left(\alpha_e, \beta_a, b_1, b_3\right)  \exp \left[-S_{a b}\left(t_a\right)\right]S_t(x_3)\right. \non
&&\left.-[2r(1-\zeta-2r_c)+r_c(1-\zeta)]\alpha_s (t_b) h\left(\alpha_e, \beta_b, b_1, b_3\right)  \exp \left[-S_{a b}\left(t_b\right)\right]S_t(r_c)\right\},
\en
where the mass ratios $r=m_D/m_{B_c}, r_b=m_b/m_{B_c}$. The hard function $h\left(\alpha_e, \beta_{a,b}, b_1, b_3\right)$, the hard scales $t_{a,b}$, the Sudakov factor $\exp \left[-S_{a b}(t)\right]$ and the threshold resummation factor $S_t(x)$ can be found in Appendix A.
The amplitudes for the nonfactorizable eimission diagrams Figs. 1(c) and 1(d) are written as
\be
\mathcal{M}_e^{L L}&=& -\frac{8}{3} \pi f_{B_c} m_{B_c}^4  C_F \int_0^1  d z d x_3 \int_0^{\infty} b_1 b d b_1 d b\exp(-\frac{\omega^2_bb^2_1}{2}) \phi_D(x_3,b_3) \phi_0(z,s) \non
&& \times\left\{\left[r\left(\zeta\left(r_c+2 z-x_3-1\right)+r_c+x_3-1\right)-r_c-z+1\right]\right. \non
&& \left.\times\alpha_s (t_c)h\left(\beta_c, \alpha_e,  b_1, b_2\right)  \exp \left[-S_{c d}\left(t_c\right)\right]\right.\non
&& \left.-\left[r\left(\zeta\left(-r_c+2 z+x_3-1\right)-r_c-x_3+1\right)\right.\right. \non
&& \left.\left.-2 \zeta\left(r_c+x_3-1\right)+2 r_c-z+x_3-1\right]\right.\non
&& \left.\times\alpha_s (t_d)h\left(\beta_d, \alpha_e, b_1, b_2\right)  \exp \left[-S_{c d}(t_d)\right]\right\},\\
\mathcal{M}_e^{L R}&=& -\frac{8}{3} \pi f_{B_c} \sqrt{\zeta} m_{B_c}^4 C_F \int_0^1  d z d x_3 \int_0^{\infty} b_1 b d b_1 d b \exp(-\frac{\omega^2_bb^2_1}{2}) \phi_D(x_3,b_1) \non
&& \times\left\{\left[r \left(\phi_s(z,s)\left(-2 r_c+z-x_3+2\right)+\zeta\left(\left(\phi_s(z,s)-\phi_t(z,s)\right)\left(x_3-z\right)\right)+\phi_t(z,s)\left(x_3-z\right)\right)\right.\right. \non
&& \left.\left.
+(\zeta-1)\left(\phi_s(z,s)+\phi_t(z,s)\right)\left(r_c+z-1\right)\right]\alpha_s (t_c)  h\left(\beta_c, \alpha_e, b_1, b\right)  \exp \left[-S_{c d}\left(t_c\right)\right]\right. \non
&&\left. -\left[r \left(\phi_s(z,s)\left(2 r_c-z+x_3-1\right)+\zeta\left(\phi_s(z,s)+\phi_t(z,s)\right)\left(-z-x_3+1\right)+\phi_t(z,s)\left(z+x_3-1\right)\right) \right.\right. \non
&& \left.\left.+(\zeta-1)\left(\phi_s(z,s)+\phi_t(z,s)\right)
\left(r_c-z\right)\right]\alpha_s (t_d)  h\left(\beta_d, \alpha_e, b_1, b\right)  \exp \left[-S_{c d}\left(t_d\right)\right]\right\},\\
\mathcal{M}_e^{S P}&=& \frac{8}{3} \pi f_{B_c} m_{B_c}^4  C_F \int_0^1  d z d x_3 \int_0^{\infty} b_1 b d b_1 d b \exp(-\frac{\omega^2_bb^2_1}{2}) \phi_D(x_3,b_1)\phi_{0}(z,b) \non
&& \times\left\{\left[r\left(\zeta\left(r_c+2 z-x_3-1\right)+r_c+x_3-1\right)+2 \zeta\left(r_c+x_3-1\right)-2 r_c-z-x_3\right]\right. \non
&& \left. \times\alpha_s (t_c)h\left(\beta_c, \alpha_e, b_1, b\right)  \exp \left[-S_{c d}\left(t_c\right)\right]\right.\non
&& \left.-\left[r\left(\zeta\left(r_c-2 z-x_3+1\right)+r_c+x_3-1\right)-r_c+z\right]\right.\non
&& \left.\times\alpha_s (t_d)h\left(\beta_d, \alpha_e, b_1, b\right)  \exp \left[-S_{c d}\left(t_d\right)\right]\right\}.
\en
It is noticed that the integration of $b_3$ has been performed using $\delta$ function $\delta(b_1-b_3)$, leaving only integration of $b_1$ and $b$.
The amplitudes from the nonfactorizable annihilation diagrams Figs. 1(e) and 1(f) are listed as
\be
\mathcal{M}_a^{L L}&=&\frac{8}{3} \pi f_{B_c} m^4_{B_c} C_F \int_0^1 d z d x_3 \int_0^{\infty} b_1 b_3 d b_1 d b_3 \exp(-\frac{\omega^2_bb^2_1}{2}) \phi_D(x_3,b_3)\non
&& \times\left\{\left[r\sqrt{\zeta} \left(\phi_s(z,s)\left(-4 r_b-2 r_c-z-x_3+2\right)+\phi_t(z,s)\left(x_3-z\right)\right.\right.\right.\non
&&\left.\left.\left.+\zeta(\phi_s(z,s)-\phi_t(z,s))\left(x_3-z\right)\right)+\phi_0(z,s)\left(r_b(\zeta-1)-r_c-z+1\right)\right]\right.\non
&&\left.\times\alpha_s (t_e) h\left(\beta_e, \alpha_a, b_1, b_3\right)  \exp \left[-S_{ef}\left(t_e\right)\right]\right.\non
&&\left.+\left[r\sqrt{\zeta} \left(\phi_s(z,s)\left(2 r_c+z+x_3\right)+\phi_t(z,s)\left(x_3-z\right)\right.\right.\right.\non
&&\left.\left.\left.\left.+\zeta\left(\phi_s(z,s)+\phi_t(z,s)\right)\left(z-x_3\right)\right)+\phi_0(z,s)\left(x_3-\zeta\left(r_c-2 z+2 x_3\right)\right)\right]\right.\right.\non
&&\left.\times\alpha_s (t_f) h\left(\beta_f, \alpha_a, b_1, b_3\right)  \exp \left[-S_{ef}\left(t_f\right)\right]\right\},
\en
\be
{M}_a^{L R}&=& \frac{8}{3} \pi f_{B_c} m_{B_c}^4  C_F   \int_0^1 d z d x_3 \int_0^{\infty} b_1 b_3 d b_1 d b_3 \exp(-\frac{\omega^2_bb^2_1}{2})\phi_D(x_3,b_3)\non
&& \times\left[r((\zeta+1)\left(-r_b+r_c-x_3-1\right)+2(z+x_3)) \phi_0(z,s)\right. \non
&& \left.+\sqrt{\zeta}(1-\zeta)\left(\phi_s(z,s)+\phi_t(z,s)\right)\left(r_b-r_c-z+1\right)\right] \non
&& \times\alpha_s (t_e)h\left(\beta_e, \alpha_a, b_1, b_3\right)  \exp \left[-S_{ef}\left(t_e\right)\right] \non
&& +\left[r((\zeta+1)\left(2r_c-x_3\right)-2(z-x_3))\phi_0(z,s)\right. \non
&& \left.+\sqrt{\zeta}(1-\zeta)\left(\phi_s(z,s)+\phi_t(z,s)\right)\left(-2r_c+z\right)\right]\non
&& \times\alpha_s (t_f) h\left(\beta_f, \alpha_a, b_1, b_3\right)  \exp \left[-S_{ef}\left(t_f\right)\right].
\en
The amplitudes from the factorizable annihilation diagrams Figs. 1(g) and 1(h) are given as
\be
\mathcal{F}_a^{L L}&=&8 \pi f_{B_c} m_{B_c}^4 C_F \int_0^1  d z d x_3 \int_0^{\infty} b b_3 d b d b_3\phi_D(x_3,b_3) \non
&&\times \left\{\left[2r\left(\zeta^{3/2}\left( x_3-1\right)-\sqrt{\zeta}\left(x_3+1\right)\right)\phi_s(z,s)+\left(\zeta\left(2 x_3-1\right)-x_3\right) \phi_0(z,s)\right]\right. \non
&&\left.\times \alpha_s (t_g) h\left(\alpha_a,\beta_g, b_2, b_3\right) \exp \left[-S_{gh}\left(t_g\right)\right]S_t(x_3)\right. \non
&&\left. -\left[2r \left(r_c(\zeta+1)\phi_0(z,s)+\zeta^{3 / 2}\left(z-1\right)\left(\phi_t(z,s)-\phi_s(z,s)\right)\right.\right.\right. \non
&& \left.-2\sqrt{\zeta}\left(\left(1+z\right) \phi_s(z,s)-\left(1-z\right) \phi_t(z,s)\right)\right)+\sqrt{\zeta}(\zeta-1) r_c\left(\phi_t(z,s)-\phi_s(z,s)\right) \non
&& \left.\left.+(\zeta-1) z \phi_0(z,s)\right]\alpha_s (t_h) h\left(\alpha_a, \beta_h, b_3, b_2\right)  \exp \left[-S_{gh}\left(t_h\right)\right]S_t(z)\right\},\\
\mathcal{F}_a^{S P}&=& 16 \pi f_{B_c} m_{B_c}^4  C_F \int_0^1  d z d x_3 \int_0^{\infty} b b_3 d b d b_3 \phi_D(x_3,b_3)\non
&&\left\{\left[-2\sqrt{\zeta} (\zeta-1)\phi_s(z,s)+r \phi_0(z,s)\left(\zeta\left(2-x_3\right)+x_3\right)\right]\right.\non
&&\left.\times\alpha_s (t_g)  h_e(\alpha_a, \beta_g, b_2, b_3)  \exp \left[-S_{gh}\left(t_g\right)\right]S_t(x_3)\right.\non
&&\left. -\left[2r\left(2\sqrt{\zeta} r_c \phi_s(z,s)-(\zeta\left(2z-1\right)+1)\phi_0(z,s)\right)\right.\right. \non
&&\left. \left.-(\zeta-1) r_c \phi_0(z,s)+\sqrt{\zeta}(\zeta-1)z(\phi_s(z,s)-\phi_t(z,s))\right]
\right.\non
&&\left.\times\alpha_s (t_h) h_e\left(\alpha_a, \beta_h, b_3, b_2\right)  \exp \left[-S_{gh}\left(t_h\right)\right]S_t(z)\right\}.
\en

By combining the amplitudes from the different Feynman diagrams, the total decay amplitudes for the quasi-two-body decays $B_c \rightarrow  K^{*}_0 D_{(s)} \rightarrow K \pi D_{(s)}$ can be written as
\be
\label{d0kstz}
\mathcal{A}\left(B_c^+ \to K^{* +}_0 D^0\to K^0\pi^+ D^0 \right)&= & \ V_{u s} V_{u b}^*\left[a_1 \mathcal{F}_e^{L L}+C_1 \mathcal{M}_e^{L L}\right]+\ V_{c s} V_{c b}^*\left[a_1 \mathcal{F}_a^{L L}+C_1 \mathcal{M}_a^{L L}\right] \non
&& -\ V_{t s} V_{t b}^*\left[\left(C_3+C_9\right)\left(\mathcal{M}_e^{L L}+\mathcal{M}_a^{L L}\right)+\left(C_5+C_7\right)\left(\mathcal{M}_e^{L R}+\mathcal{M}_a^{L R}\right)\right. \non
&& +\left(C_4+\frac{1}{3} C_3+C_{10}+\frac{1}{3} C_9\right)\left(\mathcal{F}_a^{L L}+\mathcal{F}_e^{L L}\right) \non
&& \left.+\left(C_6+\frac{1}{3} C_5+C_8+\frac{1}{3} C_7\right)\left(\mathcal{F}_a^{S P}+\mathcal{F}_e^{S P}\right)\right],\\
\mathcal{A}\left(B_c^+ \to K^{*0}_0 D^{+}\to K^+\pi^-D^+ \right)&=& V_{c s} V_{c b}^*\left[a_1 \mathcal{F}_a^{L L}+C_1 \mathcal{M}_a^{L L}\right]-V_{t s} V_{t b}^*\left[\left(C_3-\frac{1}{2} C_9\right) \mathcal{M}_e^{L L}\right. \non
&& +\left(C_3+C_9\right) \mathcal{M}_a^{L L}+\left(C_5-\frac{1}{2} C_7\right) \mathcal{M}_e^{L R}+\left(C_5+C_7\right) \mathcal{M}_a^{L R} \non
&& +\left(C_4+\frac{1}{3} C_3+C_{10}+\frac{1}{3} C_9\right) \mathcal{F}_a^{L L} \non
&&+\left(C_4+\frac{1}{3} C_3-\frac{1}{2} C_{10}-\frac{1}{6} C_9\right) \mathcal{F}_e^{L L}+\left(C_6+\frac{1}{3} C_5-\frac{1}{2} C_8\right. \non
&& \left.\left.-\frac{1}{6} C_7\right) \mathcal{F}_e^{S P}+\left(C_6+\frac{1}{3} C_5+C_8+\frac{1}{3} C_7\right) \mathcal{F}_a^{S P}\right],\label{dzkst0}\\
\mathcal{A}\left(B_c^+ \to \bar{K}^{*0}_0 D_s^{+}\to K^-\pi^+D_s^{+} \right)&= &   V_{c d} V_{c b}^*\left[a_1 \mathcal{F}_a^{L L}+C_1 \mathcal{M}_a^{L L}\right]- V_{t d} V_{t b}^*\left[\left(C_3-\frac{1}{2} C_9\right) \mathcal{M}_e^{L L}\right. \non
&& +\left(C_3+C_9\right) \mathcal{M}_a^{L L}+\left(C_5-\frac{1}{2} C_7\right) \mathcal{M}_e^{L R}+\left(C_5+C_7\right) \mathcal{M}_a^{L R} \non
&& +\left(C_4+\frac{1}{3} C_3+C_{10}+\frac{1}{3} C_9\right) \mathcal{F}_a^{L L} \non
&& +\left(C_4+\frac{1}{3} C_3-\frac{1}{2} C_{10}-\frac{1}{6} C_9\right) \mathcal{F}_e^{L L}+\left(C_6+\frac{1}{3} C_5-\frac{1}{2} C_8\right. \non
&& \left.\left.-\frac{1}{6} C_7\right) \mathcal{F}_e^{S P}+\left(C_6+\frac{1}{3} C_5+C_8+\frac{1}{3} C_7\right) \mathcal{F}_a^{S P}\right].\label{dszkst0}
\en
Using these total decay amplitudes, one can calculate the differential decay rate with the following formula
\begin{equation}
    \frac{d \mathcal{B}r}{d \omega^2}=\tau_{B_c} \frac{\left|\vec{p}_1\right|\left|\vec{p}_3\right|}{64 \pi^3 m_{B_c}^3}\left|\mathcal{A}\right|^2,
    \label{bran}
\end{equation}
where $\tau_{B_c}$ is the mean lifetime of $B_{c}$ meson, the kinematic variables
$\left|\vec{p}_1\right|$ and $\left|\vec{p}_3\right|$ denote the magnitudes of the daughter meson ($K$ or $\pi$) and the bachelor meson $D_{(s)}$ momenta in the center-of-mass frame of the $K \pi$ pair,
\begin{equation}
\begin{aligned}
\left|\vec{p}_1\right| & =\frac{1}{2} \sqrt{\left[\left(m_K^2-m_\pi^2\right)^2-2\left(m_K^2+m_\pi^2\right) w^2+w^4\right] / w^2}, \\
\left|\vec{p}_3\right| & =\frac{1}{2} \sqrt{\left[\left(m_{B_c}^2-m_D^2\right)^2-2\left(m_{B_c}^2+m_D^2\right) w^2+w^4\right] / w^2}.
\end{aligned}
\end{equation}

{\centering\section{NUMERICAL RESULTS }}
The input parameters in our numerical calculations are listed as following (the QCD scale, the masses, the decay constants and the widths are in units of $\mathrm{GeV}$, the $B_c$ meson lifetime is in units of ps) \cite{lass,pdg}
\begin{equation}
\begin{aligned}
&\Lambda_{Q C D} =0.25, m_{B_c^+}=6.274, m_b =4.8,m_{K^{\pm}}=0.494, m_{K^0}=0.498, m_{\pi^{\pm}}=0.140, m_{\pi^0} =0.135,\\
&m_{K^{*}_0(1430)}=1.425 \pm 0.050,  m_{K^{*}_0(1950)}=1.945 \pm 0.010, \Gamma_{K^{* 0}_0(1430)}=0.270 \pm 0.080,\\
& \Gamma_{K^{*}_0(1950)}=0.100\pm0.040 \;(0.201 \pm 0.034), f_{B_c}=0.489, \tau_{B_c}=0.51. \\
\end{aligned}
\end{equation}
As to the Cabibbo-Kobayashi-Maskawa (CKM) matrix elements, we employ the Wolfenstein parametrization with the inputs\cite{pdg}
\begin{equation}
\begin{aligned}
& \lambda=0.22500\pm0.00067, \quad A=0.826^{+0.018}_{-0.015}, \\
& \bar{\rho}=0.159\pm0.010, \quad \bar{\eta}=0.348\pm0.010.
\end{aligned}
\end{equation}

By using the differential branching ratio in Eq.(\ref{bran}) and the squared amplitudes in Eqs.(\ref{d0kstz})-(\ref{dszkst0}), integrating over the
full $K \pi$ invariant mass region $\left(m_K+m_\pi\right) \leq \omega \leq (M_{B_{c}}-m_{D_{(s)}})$,
we can obtain the branching ratios for these quasi-two-body decays $B_c^+ \to K^{*}_0 D_{(s)}\to K\pi D_{(s)}$ by using two different parametrizations metioned in previous section, i.e.,  the LASS line shape and the RBW function, to the intermediate
resonances. In view of the large deviations for the measurements about the $K^*_0(1950)$ decay width between the LASS \cite{lass} and BaBar \cite{babar1}, we take two width values $\Gamma_{K^*_0(1950)}=0.100\pm0.040$ GeV \cite{pdg} and $\Gamma_{K^*_0(1950)}=0.201\pm0.034$ GeV \cite{lass} in our calculations for comparison. The results are listed in Table \ref{bran12}, where the first error is from the $B_c$ meson shape parameter uncertainty $\omega_{B_c}=0.6\pm0.05$ GeV, the following two errors come
from the Gegenbauer moments in the $K \pi$ pair distribution amplitudes
$ a_{1}=-0.57 \pm 0.13, a_{2}=-0.42 \pm 0.22$, and the last one is induced by varying the
QCD scale $\Lambda_{Q C D}=0.25 \pm 0.05$ GeV from the next-to-leading-order (NLO) effect in the PQCD approach. These results are sensitive to the value of the QCD scale $\Lambda_{QCD}$, which indicate that the NLO contributions maybe play an important role in these decays. Another important uncertainty  is from the Gegenbauer coefficient
$a_{2}=-0.42 \pm 0.22$. Some comments are in order:

\begin{table}[h]
	\begin{center}
\begin{tiny}
\caption{Branching ratios of the quasi-two-body decays $B^+_c\to K^{*}_0(1430,1950)D_{(s)}\to K\pi D_{(s)}$ within the LASS and the RBW parametrizations.}
\label{bran12}
    \renewcommand\arraystretch{2}
    \begin{tabular}{cccccccc}
    \hline\hline
      Decay modes  & $\Gamma_{K^*_0}$ (GeV) &  LASS &  &RBW\\ \hline

     $B_c^+ \to \ K^{*+}(1430) D^{0}\to\ K^{0}\pi^+ D^{0}$
      & $0.270$  &  $(2.71_{-0.04-0.21-0.66-0.55}^{+0.05+0.20+0.91+1.28})\times 10^{-4} $
      &  &  $(0.61_{-0.02-0.03-0.21-0.00}^{+0.02+0.05+0.33+0.80})\times 10^{-4} $
      \\ \hline
      $B_c^+ \to \ K^{*0}(1430) D^{+}\to\ K^{+}\pi^- D^{+}$
      & $0.270$ &  $(3.06_{-0.04-0.24-0.00-0.69}^{+0.04+0.23+0.97+1.63})\times 10^{-4} $
      &  &  $(0.64_{-0.02-0.03-0.21-0.00}^{+0.02+0.05+0.33+0.88})\times 10^{-4} $
      \\ \hline
    $B_c^+ \to \bar K^{*0}(1430) D_s^{+} \to K^{-}\pi^+ D_s^{+}$
      & $0.270$ &  $(2.04_{-0.03-0.15-0.47-0.45}^{+0.04+0.16+0.62+1.10})\times 10^{-5}  $
      &  &  $(0.48_{-0.01-0.04-0.16-0.00}^{+0.01+0.04+0.25+0.68})\times 10^{-5}  $
      \\  \hline\hline
      $B_c^+ \to \ K^{*+}(1950) D^{0}\to\ K^{0}\pi^+ D^{0}$         & $0.100$     &  $(7.76_{-0.12-0.19-2.41-1.30}^{+0.12+0.22+3.11+1.63})\times 10^{-5} $
                                                                                      &                       &  $(2.73_{-0.05-0.04-1.00-2.36}^{+0.04+0.02+1.19+1.91})\times 10^{-5} $\\
                                                                                      & $0.201$     &  $(1.99_{-0.03-0.06-0.65-0.89}^{+0.03+0.06+0.84+0.19})\times 10^{-5} $
                                                                                      &                       &  $(1.31_{-0.02-0.02-0.48-1.03}^{+0.02+0.00+0.57+0.77})\times 10^{-5} $      \\ \hline
    $B_c^+ \to \ K^{*0}(1950) D^{+}\to\ K^{+}\pi^- D^{+}$         & $0.100$   &  $(7.80_{-0.14-0.16-2.38-0.54}^{+0.14+0.23+3.39+2.35})\times 10^{-5} $
                                                                                      &                       &  $(2.55_{-0.05-0.02-0.94-2.18}^{+0.06+0.04+1.42+2.13})\times 10^{-5} $\\
                                                                                      & $0.201$     &  $(2.01_{-0.04-0.06-0.63-0.05}^{+0.04+0.08+0.92+1.10})\times 10^{-5} $
                                                                                      &                       &  $(1.27_{-0.03-0.00-0.46-0.98}^{+0.03+0.01+0.66+0.92})\times 10^{-5} $\\ \hline
      $B_c^+ \to \bar K^{*0}(1950) D_s^{+} \to K^{-}\pi^+ D_s^{+}$  & $0.100$    &  $(5.47_{-0.12-0.05-1.59-0.41}^{+0.12+0.16+2.21+1.79})\times 10^{-6} $
                                                                                      &                       &  $(1.82_{-0.05-0.00-0.62-1.54}^{+0.04+0.72+0.97+1.52})\times 10^{-6} $\\
                                                                                      & $0.201$    &  $(1.37_{-0.03-0.02-0.41-0.03}^{+0.03+0.06+0.59+0.83})\times 10^{-6} $
                                                                                      &                       &  $(0.91_{-0.02-0.00-0.30-0.70}^{+0.02+0.03+0.44+0.67})\times 10^{-6} $\\ \hline\hline
 \end{tabular}
 \end{tiny}
 \end{center}
\end{table}
\begin{enumerate}

\item
    For the decays with $K^*_0(1430)$ involved there exists a significant difference in the branching ratios obtained using these two different parameterization schemes to the time-like from factor. If taken the RBW model with only the resonance contribution being considered, the corresponding branching ratios will be much smaller than those obtained using the LASS form, where both the resonance and nonresonance contributions are included. For example, $\mathcal{B}r(B_c^+ \to \bar K^{*0}_0(1430) D_s^{+} \to K^{-}\pi^+ D_s^{+})$ under the RBW parameterization is only one fifth of that under the LASS form. Such large difference should be detected by the future experiments, then one can clarify
    which parameterization scheme is more reasonable. The branching ratios of the decays with $D^{0,+}$ mesons involved are about one order larger than that of the decay with $D^+_s$ involved, this is mainly because that the CKM element $V_{cs}(V_{ts})$ associated with the former is about 5 times of $V_{cd}(V_{td})$ associated with the latter. For the decay $B_c^+ \to \ K^{*+}_0(1430) D^{0}\to\ K^{0}\pi^+ D^{0}$, the main contributions come from the annihilation type amplitudes with the tree operators $F^{LL}_a$ and $M^{LL}_a$. Such contributions are much larger than another kind of tree contributions associated with $V_{us}V^*_{ub}$. Although the values of the CKM matrix elements $V^*_{cb}V_{cs}$ and $V^*_{tb}V_{ts}$ are close to each other, the contributions from the penguin operators are small and only about $2\%$ because of the smallness of the Wilson coefficients. These characters are very different with the case of the decay $B_c^+ \to \ K^{*+}(892) D^{0}\to\ K^{0}\pi^+ D^{0}$ \cite{zyzhang}.
\item
    As we know, the narrow resonances, such as $K^*(892)$, can be well described by the RBW model, which has been adopted extensively in experimental analysis. While it is failed
    to describe the broad resonances, such as $K^*_0(1430)$, which is usually considered to be interfer strongly with a slowly varying nonresonant contribution \cite{meadows}. On the experimental side, the LASS line shape was proposed to describe the combined contributions from the resonant and non-resonant components \cite{lass}. As to the decays with $K^*_0(1430)$ involved, the resonant contributions come from the second term in the LASS parameterization shown in Eq. (\ref{lass}) are only about $20\%$ of the total branching ratios. It is similar to the case of the decays $B\to KKK $, where the nonresonant background are dominant. For example, the resonant fraction is about $10\%$ in the channel $B^0\to K^+K^-K^0$ \cite{BaBar2007}. In the decays $B\to K^*_0(1430)D\to K\pi D$, the resonant contributions are estimated to be $50\%$ \cite{wsfang}.
\item
For the decays with $K^*_0(1950)$ involved, besides two parametrization schemes we use two width values $\Gamma_{K^*_0(1950)}=0.100, 0.201$ GeV in our calculations. In Table \ref{bran12}, one can find that the differences of the branching ratios between these two parameterization schemes are not very large for $\Gamma_{K^*_0(1950)}=0.201$ GeV. The branching ratios obtained in the LASS (RBW) parametrization are (not) sensitive to the values of $\Gamma_{K^*_0(1950)}$.
On the other hand, one can find that the proportions of the resonant contributions in the decay channels containing $K^*_0(1950)$ are larger compared to those in the decays
    with $K^*_0(1430)$ involved. The resonant contributions are more than $30\% (60\%)$ with $\Gamma_{K^*_0(1950)}=0.100 (0.201)$ GeV. In addition, there exist destructive interferences with different strengths between the resonant and nonresonant contributions in these considered decays $B_c\to K^*_0(1430, 1950) D_{(s)}$.
\end{enumerate}

Under the so-called narrow width approximation (NWA) one can extract the branching fraction of the two-body decay $B_c\to RP$ from that of the quasi-two-body one $B_c\to RP\to P_1P_2P$, that is
\begin{equation}
\Gamma\left(B_c \rightarrow R P \rightarrow P_1 P_2 P\right)=\Gamma\left(B_c \rightarrow R P\right) \mathcal{B}r\left(R \rightarrow P_1 P_2\right).
\label{nwa}
\end{equation}
Here the strong decay $R \to P_1 P_2$ represents $K^*_0(1430)\to K\pi$ and $K^*_0(1950)\to K\pi$, whose branching ratios are given as $0.93\pm0.10$ and $0.52\pm0.14$ \cite{pdg}, respectively, and the branching ratios of the decays $K^{* 0}_0(1430,1950) \rightarrow K^{+} \pi^{-}$ and $K^{* +}_0(1430,1950) \rightarrow K^{0} \pi^{+}$can be obtained from the isospin conservation, namely
\begin{equation}
\begin{gathered}
\frac{\Gamma\left(K^{* 0}_0(1430,1950) \rightarrow K^{+} \pi^{-}\right)}{\Gamma\left(K^{* 0}_0(1430,1950) \rightarrow K \pi\right)}=
\frac{\Gamma\left(K^{*+}_0(1430,1950) \rightarrow K^0 \pi^{+}\right)}{\Gamma\left(K^{*+}_0(1430,1950) \rightarrow K \pi\right)}=2 / 3.
\end{gathered}
\end{equation}
In Eq. (\ref{nwa}), $R$ refers to an intermediate resonant state with the zero width limit $\Gamma_R\to 0$. While the scalar resonances $K^*_0(1430,1950)$ considered here have broad decay widths, one should take into account the finite-width effects, which are defined as $\eta_R$. Here we take $\eta_{K_0^*(1430)}$ within the RBW parameterization as an example,
\begin{equation}
\begin{aligned}
\eta_{K_0^*(1430)}& =\frac{\Gamma\left(B_c \rightarrow K_0^*(1430) D_{(s)} \rightarrow K \pi D_{(s)}\right)}{\Gamma\left(B_c \rightarrow K_0^*(1430) D_{(s)}\right) \times \mathcal{B}\left(K_0^* \rightarrow K \pi\right)} \\
& \approx \frac{m_{K_0^*(1430)}^2}{4 \pi m_{B_c}} \frac{\Gamma_{K_0^*(1430)}}{q_{D_{(s)}} q_0} \int_{\left(m_K+m_\pi\right)^2}^{\left(m_{B_c}-m_{D_{(s)}}\right)^2} \frac{d s}{s} \frac{\lambda^{1 / 2}\left(m^2_{B_c}, s, m_{D_{(s)}}^2\right) \lambda^{1 / 2}\left(s, m_K^2, m_\pi^2\right)}{\left(s-m_{K_0^*(1430)}^2\right)^2+\left(m_{K_0^*(1430)} \Gamma_{K_0^*(1430)}(s)\right)^2},
\end{aligned}
\end{equation}
where $\lambda(a, b, c)=a^2+b^2+c^2-2 a b-2 a c-2 b c$, and $q_{D_{(s)}}$ refers to the magnitude momentum of the bachelor meson $D_{(s)}$ and is given as
\be
q_{D_{(s)}}=\frac{1}{2} \sqrt{\left[\left(m^2_{B_c}-m_{D_{(s)}}^2\right)^2-2\left(m_{B_c}^2+m_{D_{(s)}}^2\right) s+s^2\right] / s},
\en
with $s$ being fixed at $m_{K_0^*(1430)}^2$. For our considered decays, the values of $\eta_R$ are calculated as $\eta_{K_0^*(1430)}=0.90$ and $\eta_{K_0^*(1950)}=0.94(0.97)$
with $\Gamma_{K_0^*(1950)}=0.201(0.100)$ GeV. As to $\eta_{K_0^*(1430)}$, it was also discussed in Ref. \cite{cheng}, where its values were given as $\eta^{\text{QCDF}}_{K_0^*(1430)}=0.83$ and $\eta^{\text{EXPP}}_{K_0^*(1430)}=1.10$ \footnote{The definition of $\eta_{R}$ and that used in Ref.\cite{cheng} are the inverse of each other. }. The former was calculated within the QCD factorization (QCDF) framework, and the latter was abstracted from the data.
It is obviously that our value of $\eta_{K_0^*(1430)}$ is larger than the QCDF result, but smaller than that abstracted from the data.
After considering the decay width effects, the branching ratios for the two-body decays $B_c\to K^*_0D_{(s)}$ can be related to our considered decays from the following
formula
\be
Br\left(B_c \rightarrow \ K^{*}_0 D_{(s)} \rightarrow K\pi D_{(s)} \right)
= \eta_{K^*_0} \cdot Br\left(B_c \rightarrow  \ K^{*}_0 D_{(s)}\right) \cdot Br\left(K^{*}_0 \rightarrow K \pi\right).
\en
\begin{table}[H]
	\caption{The branching ratios of the (quasi-)two-body decays $B_c \to  K^{*}_0(1430)D_{(s)}(\to K\pi D_{(s)})$ within the LASS and the RBW parametrization schemes. For comparison, we also list the previous PQCD calculations in the two-body framework under SII as given in Ref.\cite{Zou:2017yxc}.}
	\centering
	\renewcommand\arraystretch{2}
	\begin{tabular}{ccccc}
		\hline
		Decay modes                                      &  scheme  &  Quasi-two-body                       & Two-body         & PQCD (SII)\cite{Zou:2017yxc}   \\ \hline
		$B_c^{+} \to K^{*+}_0(1430)D^0(\to K^0\pi^+D^0)$  &LASS  &$(2.71_{-0.89}^{+1.58})\times 10^{-4}$  &$(3.97_{-1.30}^{+2.32})\times 10^{-4}$   & $(4.58\pm2.42)\times 10^{-4}$       \\
		&RBW   &$(0.61_{-0.21}^{+0.94})\times 10^{-4}$  &$(0.89_{-0.31}^{+1.37})\times 10^{-4}$    &                            \\\hline
		$B_c^{+} \to K^{* 0}_0(1430) D^{+}(\to K^+\pi^- D^+)$ &LASS  &$(3.06_{-0.73}^{+1.91})\times 10^{-4}$  &$4.48_{-1.07}^{+2.80})\times 10^{-4}$   & $(4.81^{+2.44}_{-2.73})\times 10^{-4}$       \\
		&RBW   &$(0.64_{-0.22}^{+0.94})\times 10^{-4}$  &$(0.94_{-0.32}^{+1.38})\times 10^{-4}$   &                              \\\hline
		$B_c^{+} \to \bar{K}^{* 0}_0(1430)D_s^{+}(\to K^-\pi^+D^+_s)$ &LASS  &$(2.04_{-0.67}^{+1.27})\times 10^{-5}$  &$(2.99_{-0.98}^{+1.86})\times 10^{-5}$   & $(2.79^{+1.70}_{-1.31})\times 10^{-5}$       \\
		&RBW   &$(0.48_{-0.17}^{+0.73})\times 10^{-5}$  &$(0.70_{-0.25}^{+1.07})\times 10^{-5}$   &                              \\\hline
		\hline
	\end{tabular}
	\label{tabbran2}
\end{table}
\begin{table}[H]
	\begin{footnotesize}
		\caption{Same as Table \ref{tabbran2} except for the (quasi-)two-body decays $B_c \to  K^{*}_0(1950)D_{(s)}(\to K\pi D_{(s)})$.  }
		\centering
		\renewcommand\arraystretch{2}
		\begin{tabular}{cccccc}
			\hline\hline
			Decay Modes                                 &    $\Gamma_{K^*_0(1950)}$                &  Quasi-two-body (LASS)                               & Two-body(LASS)       &  Quasi-two-body(RBW)                           & Two-body(RBW)\\
			\hline
			
			$B_c^{+} \to K^{*+}_0(1950) D^0$               &$0.100$GeV  &  $7.76_{-2.74}^{+3.52} \times 10^{-5}$       &  $1.77_{-0.63}^{+0.80}\times 10^{-4}$     &$2.73_{-2.56}^{+2.25} \times 10^{-5}$    &$5.42_{-5.08}^{+4.47}\times 10^{-5}$\\
$(\to K^0\pi^+ D^0)$                           &$0.201$GeV  &  $1.99_{-1.10}^{+0.86} \times 10^{-5}$       &  $4.54_{-2.46}^{+1.92}\times 10^{-5}$     &$1.31_{-1.14}^{+0.96} \times 10^{-5}$    &$2.99_{-2.60}^{+2.19}\times 10^{-5}$\\
\hline
$B_c^{+} \to K^{* 0}_0(1950) D^{+}$            &$0.100$GeV  &  $7.80_{-2.45}^{+4.13} \times 10^{-5}$       &  $1.78_{-0.56}^{+0.94}\times 10^{-4}$     &$2.55_{-2.37}^{+2.56} \times 10^{-5}$    &$5.82_{-5.41}^{+5.84}\times 10^{-5}$\\
$(\to K^+\pi^- D^+)$                           &$0.201$GeV  &  $2.01_{-0.63}^{+0.93} \times 10^{-5}$       &  $4.59_{-1.44}^{+2.12}\times 10^{-5}$     &$1.27_{-1.08}^{+1.13} \times 10^{-5}$    &$2.90_{-2.47}^{+2.58}\times 10^{-5}$\\
\hline
$B_c^{+} \to \bar{K}^{* 0}_0(1950) D_s^{+}$    &$0.100$GeV  &  $5.47_{-1.65}^{+2.85} \times 10^{-6}$       &  $1.25_{-0.38}^{+0.65}\times 10^{-5}$     &$1.82_{-1.66}^{+1.94} \times 10^{-6}$    &$4.16_{-3.79}^{+4.43}\times 10^{-6}$\\
$(\to K^-\pi^+ D^+_s)$                         &$0.201$GeV  &  $1.37_{-0.41}^{+1.02} \times 10^{-6}$       &  $3.13_{-0.94}^{+2.33}\times 10^{-6}$     &$0.91_{-0.76}^{+0.80} \times 10^{-6}$    &$2.08_{-1.74}^{+1.83}\times 10^{-6}$\\
\hline\hline
		\end{tabular}
		\label{tabbran3}
	\end{footnotesize}
\end{table}
The corresponding results are listed in Tables \ref{tabbran2} and \ref{tabbran3}, where the errors are the same with those given in Table \ref{bran12}. In Table \ref{tabbran2}, we also present the previous PQCD results calculated in the two-body framework \cite{Zou:2017yxc}, where $K^*_0(1430)$ is considered as the lowest lying $q\bar s$ state (SII) and the first excited $q\bar s$ state (SI), respectively. The branching ratios of the two-body decays $B_c\to  K^{*}_0(1430)D_{(s)}$ obtained from the NWA with the decay width effects considered under the LASS parameterization are consistent well with the previous PQCD results calculated in SII, while the branching ratios under the RBW parameterization are much smaller than the previous results given in the two scenarios. The scalar meson $K^*_0(1430)$ has also been studied in the $B_{(s)}$ decays by many authors \cite{shen,Cheng:2013fba}, where the SII explanation is more favored. If the decays $B_c\to K^*_0(1430)D_{(s)}$ can be detected by the future experiments with the large branching ratios at the order of $10^{-5}\sim10^{-4}$, one can speculate that the LASS parameterization is more preferable to describe the $K^*_0(1430)$.

Now we turn to the evaluations of the direct CP violation for the these considered decays, which are induced by the interference between the tree and penguin amplitudes and can be defined as
\begin{equation}
    A_{CP}=\frac{\Gamma\left(B_c^- \to \bar f \right)-\Gamma\left(B_c^+ \to f \right)}{\Gamma\left(B_c^- \to \bar f \right)
    +\Gamma\left(B_c^+ \to f \right)}.
    \label{cpeq}
\end{equation}
Here $\bar f$ is the CP conjugated final state of $f$. The numerical results are listed in Table \ref{tabcp}. It is obvious that these results are not sensitive to the parameters in the DAs, but suffer from large uncertainties due to the QCD
scale $\Lambda_{QCD}$, which can be reduced by including the high order corrections. From the numerical results, we find the following points:
\begin{table}[H]
\begin{footnotesize}
\caption{The direct CP violations ($\%$) of the decays $B_c\to K^*_0D_{(s)}(\to K\pi D_{(s)})$, where the errors are the same with those given in Table \ref{bran12}. The previous PQCD results (SII) \cite{Zou:2017yxc} within the two-body frmaework are also lited for comparison. }
    \centering
    \renewcommand\arraystretch{2}
    \begin{tabular}{ccccccc}
    \hline\hline
          Decay modes  &  &  LASS &  & RBW & & PQCD (SII) \cite{Zou:2017yxc} \\ \hline
            $B_c^+ \to \ K^{*+}(1430) D^{0}(\to\ K^{0}\pi^+ D^{0})$
             &  &  $0.58_{-0.01-0.32-0.71-2.16}^{+0.01+0.36+0.86+0.65} $
             &  &  $-1.84_{-0.22-0.67-0.59-0.74}^{+0.23+0.71+1.39+2.83}$
             &  & $0.24^{+0.14}_{-0.14}$
             \\ \hline
             $B_c^+ \to \ K^{*0}(1430) D^{+}(\to\ K^{+}\pi^- D^{+})$
             &  &  $ 0.09_{-0.00-0.01-0.04-0.09}^{+0.00+0.02+0.03+0.14} $
             &  &  $0.29_{-0.01-0.04-0.09-0.34}^{+0.01+0.03+0.02+0.02} $
             &  &  $0.0$
             \\ \hline
             $B_c^+ \to \bar K^{*0}(1430) D_s^{+} (\to K^{-}\pi^+ D_s^{+})$
             &  &  $ -5.48_{-0.31-0.37-0.00-2.46}^{+0.31+0.48+0.25+1.93} $
             &  &  $-9.71_{-0.58-0.94-0.00-0.00}^{+0.58+1.16+0.72+6.61} $
             &  &  $-3.14^{+1.72}_{-2.17}$
             \\ \hline
             $B_c^+ \to \ K^{*+}(1950) D^{0}(\to\ K^{0}\pi^+ D^{0})$
             &  &  $-1.29_{-0.12-0.35-0.22-0.48}^{+0.12+0.41+0.38+0.16} $
             &  &  $-2.40_{-0.20-0.45-0.19-2.14}^{+0.19+0.50+0.17+0.91} $
             \\ \hline
             $B_c^+ \to \ K^{*0}(1950) D^{+}(\to\ K^{+}\pi^- D^{+})$
             &  &  $0.24_{-0.01-0.04-0.01-0.15}^{+0.01+0.04+0.00+0.00} $
             &  &  $ 0.37_{-0.01-0.06-0.05-0.15}^{+0.01+0.05+0.06+0.22}$
             \\ \hline
             $B_c^+ \to \bar K^{*0}(1950) D_s^{+} (\to K^{-}\pi^+ D_s^{+})$
             &  &  $ -6.37_{-0.26-0.88-0.46-0.00}^{+0.26+1.03+0.58+1.89} $
             &  &  $-7.26_{-0.20-1.17-1.07-7.19}^{+0.21+1.42+1.10+2.15} $\\\hline\hline
 \end{tabular}
\label{tabcp}
\end{footnotesize}
\end{table}

\begin{enumerate}
\item
     The direct CP violations for these considered decays predicted by using the LASS parametrization are closer to the
     previous PQCD calculations under the two-body framework compared to those obtained under the RBW form. It indicates that the LASS
     parametrization is more suitable to descibe the scalar meson $K^*_0(1430)$. Especially for the decay $B^+_c\to K^{*+}_0(1430)D^0\to K^0\pi^+D^0$, its direct CP violation is minus under the RBW parametrization, while the sign will be flipped by the contribution from the nonresonant compoent in the LASS form.

\item
    As to the decays with the scalar meson $K^*_0(1950)$ involved, their direct CP violations are similar to those of the corresponding channels with $K^*_0(1430)$ involved. This is consistent with our expectation that these two scalar mesons should have simliar properties in the $B_c$ meson decays, while there seem to exist some differences between these two mesons: the CP violations for the decays involving $K^*_0(1950)$ between these two parametrizations are close to each other, indicating that the effect from the nonresonant contribution may not be so serious as that for the case of decays involving $K^*_0(1430)$. 
\item
    The amplitudes of the decay $B_c^{+} \rightarrow \bar{K}^{* 0}_0 D_s^{+} \rightarrow K^{-} \pi^{+} D_s^{+}$ can be obtained from those of the channel $B_c^{+} \rightarrow K^{* 0}_0 D^{+} \rightarrow K^{+} \pi^{-} D^{+}$ by replacing $D^{+}\left(V_{t s}, V_{c s}\right)$ with $D_s^{+}\left(V_{t d}, V_{c d}\right)$. The total decay amplitudes for these two decays can be rewritten as
     \begin{equation}
     \mathcal{A}=V_{c b}^* V_{c q} T-V_{t b}^* V_{t q} P=V_{c b}^* V_{c q} T\left[1+z e^{i(\alpha+\delta)}\right],
     \end{equation}
where $T$ and $P$ are the tree and penguin amplitudes, $\alpha$ and $\delta$ are the weak and strong phases, respectively. The parameters $z$ and $\alpha$ are defined as
     \begin{equation}
     z=\left|\frac{V_{t b}^* V_{t q}}{V_{c b}^* V_{c q}} \frac{P}{T}\right|, \quad \alpha=\arg \left[-\frac{V_{t b}^* V_{t q}}{V_{c b}^* V_{c q}}\right],
     \end{equation}

    with $q=d(s)$ for the decay $B_c^{+} \rightarrow \bar{K}^{* 0}_0 D_s^{+} \rightarrow K^{-} \pi^{+} D_s^{+}\left(B_c^{+} \rightarrow K^{* 0}_0 D^{+} \rightarrow K^{+} \pi^{-} D^{+}\right)$. Then the direct CP asymmetry is written as
     \begin{equation}
     A_{C P}=\frac{2 z \sin \alpha \sin \delta}{z^2+1+2 \cos \alpha \cos \delta} .
     \end{equation}

    As the weak phases are measured as $\arg \left[-\frac{V_{t b}^* V_{t d}}{V_{c b}^* V_{c d}}\right] \sim-0.40$ and $\arg \left[-\frac{V_{t b}^* V_{t s}}{V_{c b}^* V_{c s}}\right] \sim 0.02$ \cite{pdg}, their corresponding sine values are about -0.39 and 0.02 . So one can find that the size of $A_{C P}\left(B_c \rightarrow \bar{K}^{* 0}_0 D_s^{+}\rightarrow K^{-} \pi^{+} D_s^{+}\right)$ is larger than that of $A_{C P}\left(B_c^{+} \rightarrow K^{* 0}_0 D^{+}\rightarrow K^{+} \pi^{-} D^{+}\right)$ because of the larger weak phase. It is similar to the case with the S-wave resonance $K^*_0$ replaced by the P-wave resonance $K^*(892)$ in these decays \cite{zyzhang}. But the sine values of the strong phases for these two kinds of decays have opposite signs. That is to say, $\sin\delta$ for the considered decays with the S-wave resonance $K^*_0$ involved is positive, while that for the decays with the P-wave resonance $K^*(892)$ involved is negtive.
\end{enumerate}

\begin{figure}[H]
\centering
    \begin{minipage}[t]{0.497\textwidth}
        \centering
        \includegraphics[width=1\textwidth]{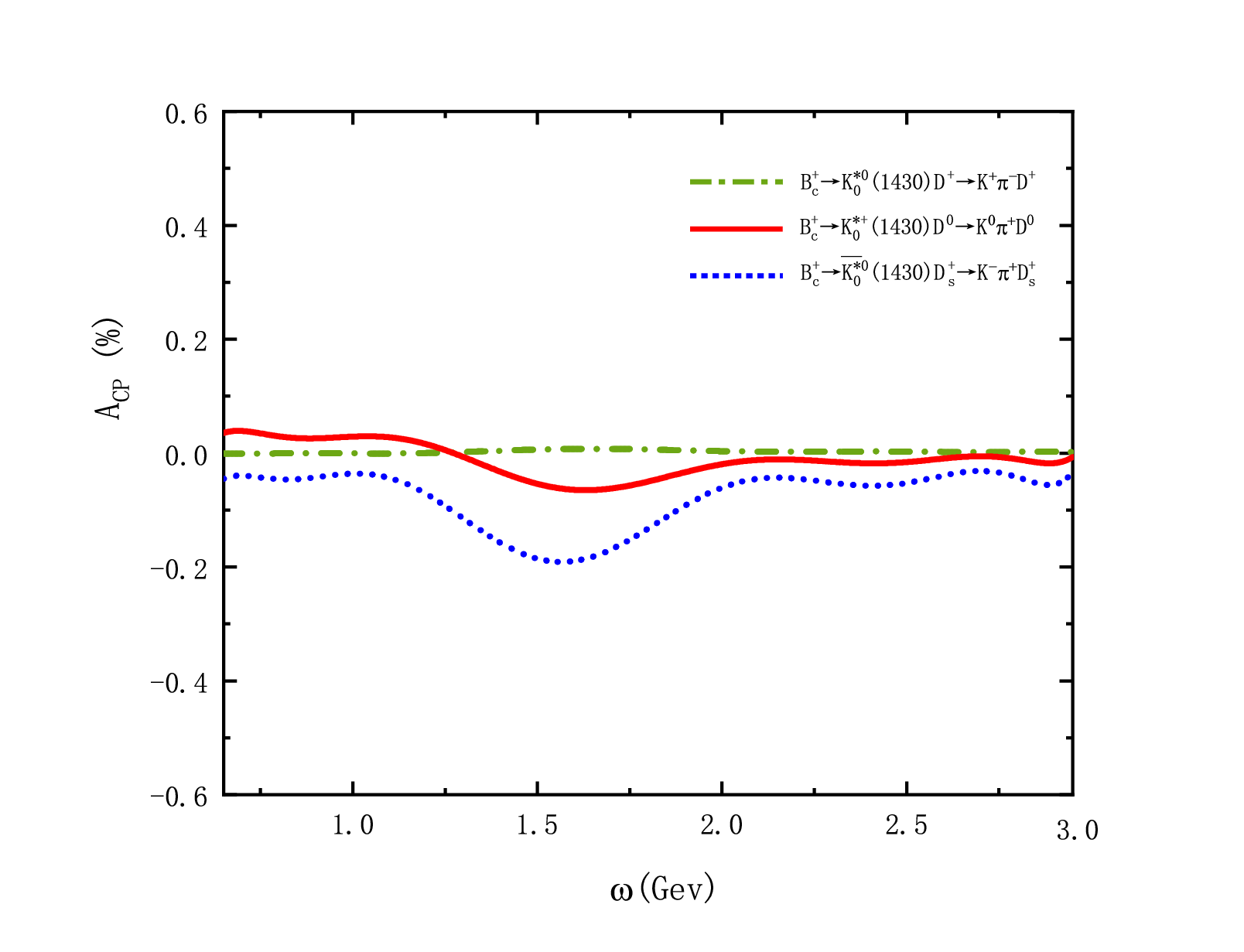}
        \centerline{(a)}
    \end{minipage}
    \begin{minipage}[t]{0.497\textwidth}
         \centering
         \includegraphics[width=1\textwidth]{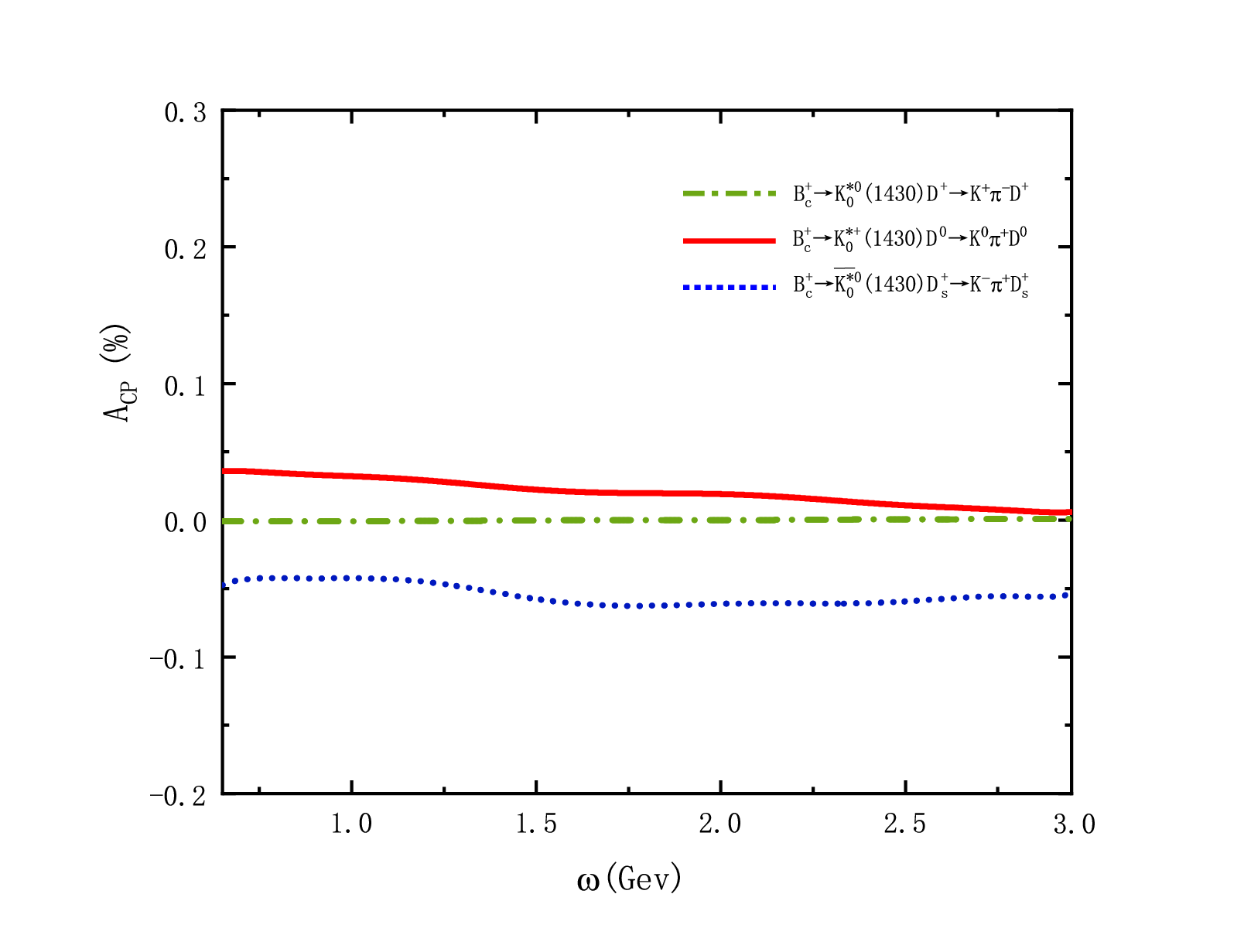}
         \centerline{(b)}
    \end{minipage}
    \caption{ The differential distributions of $A_{CP}$ in $\omega$ for the decays $B_c\to K^{*}_0(1430) D_{(s)}  \to K \pi  D_{(s)} $ obtained in the
    	BRW (left) and the LASS (right) parametrizations  . }
        \label{cp1430}
   \end{figure}

\begin{figure}[H]
\centering
    \begin{minipage}[t]{0.497\textwidth}
        \centering
        \includegraphics[width=1\textwidth]{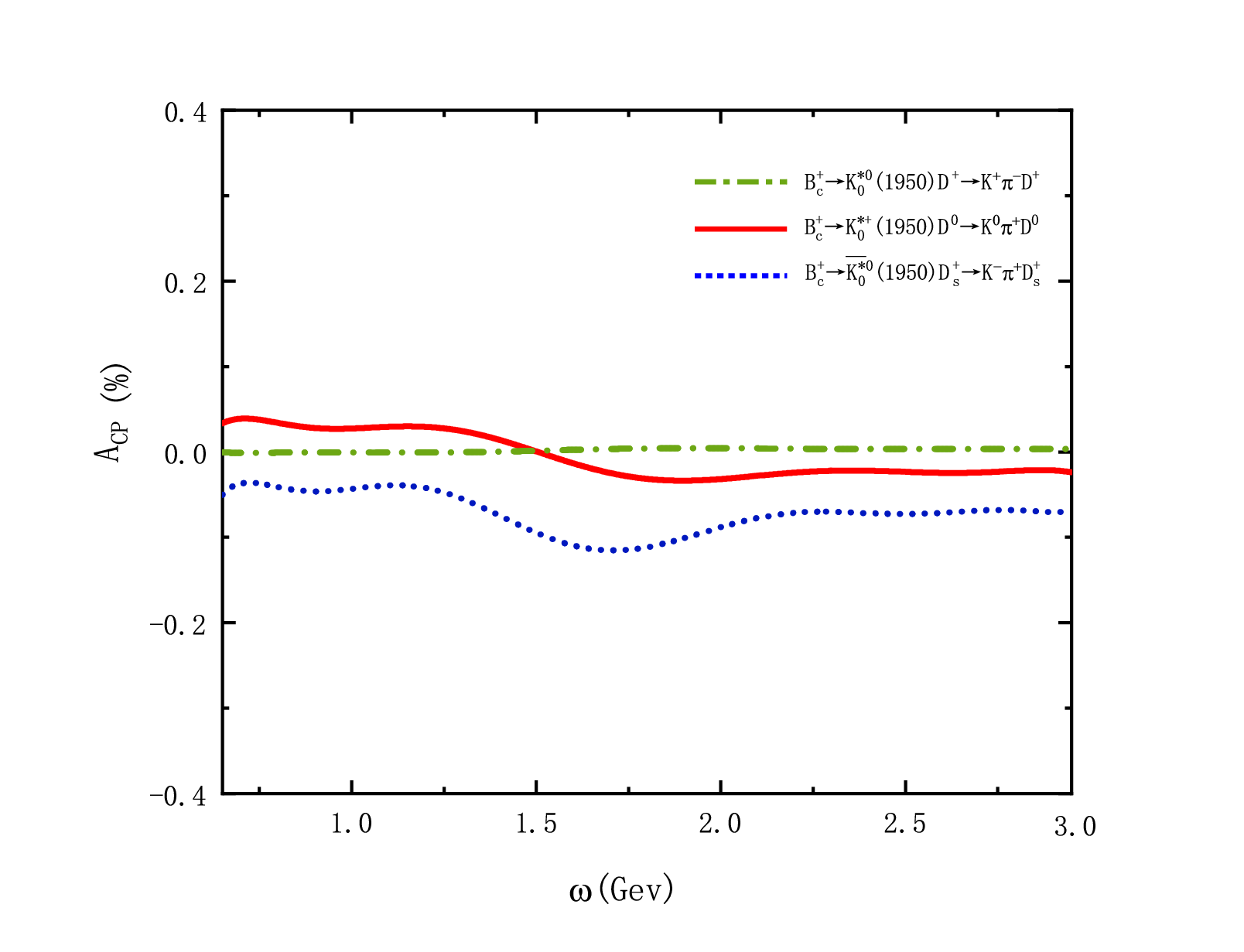}
        \centerline{(a)}
    \end{minipage}
    \begin{minipage}[t]{0.497\textwidth}
         \centering
         \includegraphics[width=1\textwidth]{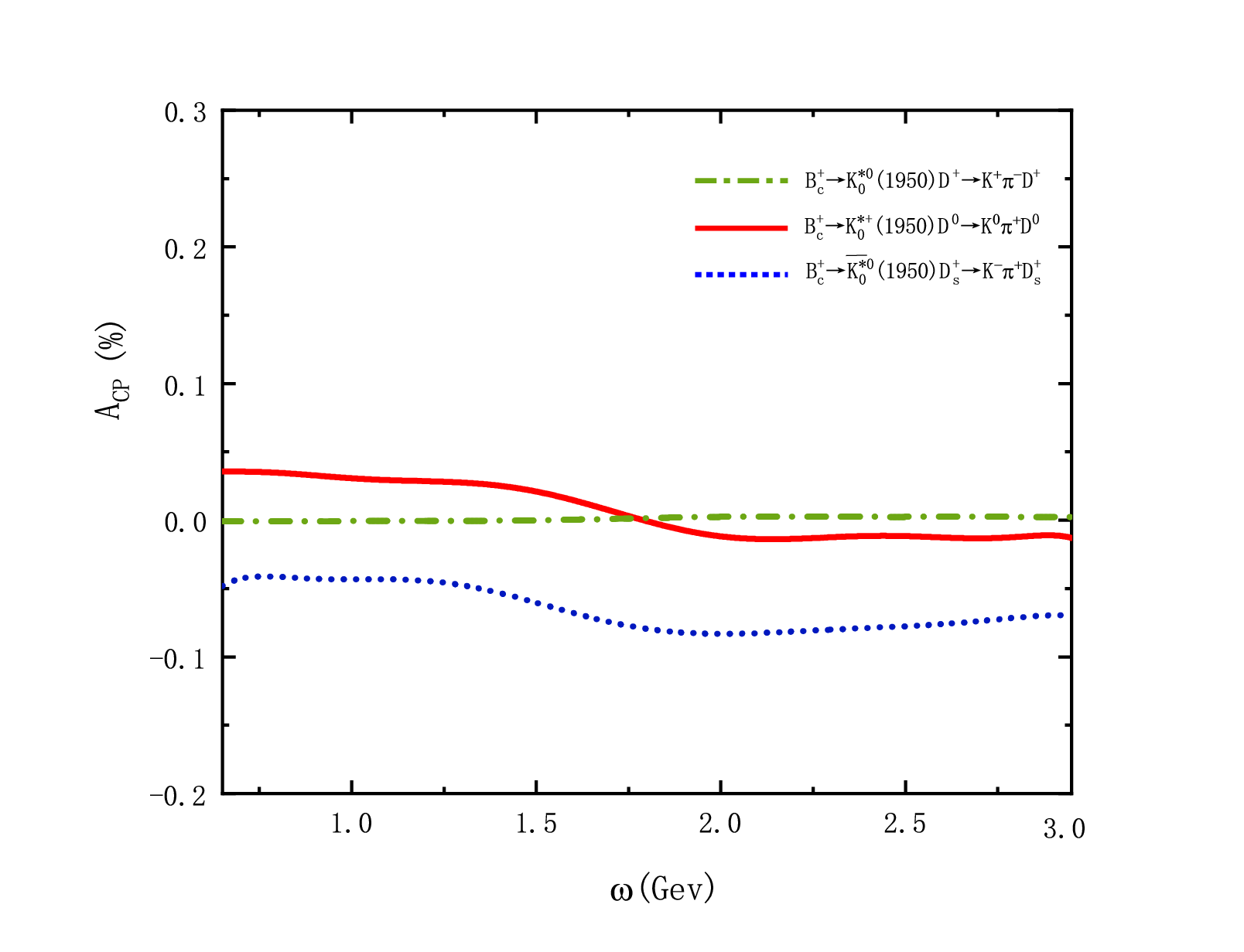}
         \centerline{(b)}
    \end{minipage}
    \caption{ The differential distributions of $A_{CP}$ in $\omega$ for the decays $B_c\to K^{*}_0(1950) D_{(s)}  \to K \pi  D_{(s)} $ obtained in the
    	BRW (left) and the LASS (right) parametrizations. }
        \label{cp1950}
   \end{figure}

The direct CP violation is just a number in the two-body framework, where the
resonance mass is fixed during the calculations. While the CP violation under the three-body framework is dependent on the $K \pi$ invariant mass $\omega$. We plot the differential distributions of the direct CP violations for the considered decays in Figures \ref{cp1430} and \ref{cp1950}. In Figure \ref{cp1430}, we can find that the differential distributions of the $A_{CP}$ for the decay $B_c^+\to \bar K^{*0}_0(1430)D^+_s\to K^-\pi^+D^+_s$ in both the BRW and LASS parametrizations are negative, those for the decay
$B_c^+ \to K^{*0}_0(1430) D^{+}\to\ K^{+}\pi^- D^{+}$ are close to zero in the whole $K \pi$ invariant mass region. The differential distribution of the $A_{CP}$ for the decay
$B_c^+ \to K^{*+}_0(1430) D^{0}\to\ K^{0}\pi^+ D^{0}$ undergoes a sign flip in the BRW parametrization, while it is positive in the LASS form. In Figure \ref{cp1950}, one can find that the differential distributions of the $A_{CP}$ for the decays involving the $K^*_0(1950)$ in both the BRW and the LASS parametrizations are similar with each other, which might also indicate that the effect from nonresonant component in the $K^*_0(1950)$ is not as serious as that in the $K^*_0(1430)$.


{\centering\section{SUMMARY}}

In this paper, we studied the quasi-two-body decays
$B_c\to K^{*}_0(1430,1950) D_{(s)}\to K\pi D_{(s)} $ using the PQCD approach. The S-wave $K \pi$ pair DAs were introduced to discribe the final-state interactions between $K$ and $\pi$ in the resonant region. As the crucial nonperturbative input, the time-like form factor $F_{K \pi}$ was parameterized by using two kinds of parametrizations, i.e., the LASS line shape and RBW model. Under the narrow width approximation and the isospin conservation, the branching ratios of these quai-two-body decays can be related with those of the corresponding two-body decays $B_c \to K^{*}_0(1430,1950) D_{(s)}$. We found that taking the LASS parametration, both the
branching ratios and the direct CP violations for the decays $B_c\to K^{*}_0(1430) D_{(s)}$ can be consistent with the previous PQCD calculations under the two-body framework by asumming $K^*_0(1430)$ as the lowest lying $q\bar s$ state (SII). Under the LASS parametration, a slowly varying non-resonant component which intefers strongly with the resonance gives the dominant contribution to the large branching ratios with $10^{-5}\sim10^{-4}$ order. As to the decays with $K^{*}_0(1950)$ involved, besides the two kinds of parametrizations,  two decay width values $\Gamma_{K_0^*(1950)}=0.201$ GeV and $\Gamma_{K_0^*(1950)}=0.100$ GeV were used in the calculations. When taking $\Gamma_{K_0^*(1950)}=0.201$ GeV, one can find that the branching ratios for the decays $B_c\to K^{*}_0(1950) D_{(s)}$ between these two parametrization schemes are close to each other, which are about one order smaller than those of the corresponding decays $B_c\to K^{*}_0(1430) D_{(s)}$. The effect from the non-resonant contribution in the decays with $K^{*}_0(1950)$ involved is not as serious as that in the channels with  $K^{*}_0(1430)$ involved. These results can be tested by the future experiments.

\section*{Acknowledgment}
This work is partly supported by the National Natural Science
Foundation of China under Grant No. 11347030, by the Program of
Science and Technology Innovation Talents in Universities of Henan
Province 14HASTIT037, Natural Science Foundation of Henan
Province under Grant No. 232300420116.


{\centering\section{Appendix A: Some relevant functions}}

We show the explicit expressions of the hard functions $h_i$ with $i=(a, \cdots, h)$, which are obtained from the Fourier transform of the hard kernels
\begin{equation}
\begin{aligned}
h_i\left(\alpha, \beta, b_1, b_2\right) & =h_1\left(\beta, b_2\right) \times h_2\left(\alpha, b_1, b_2\right), \\
h_1\left(\beta, b_2\right) & = \begin{cases}K_0\left(\sqrt{\beta} b_2\right), & \beta>0, \\
K_0\left(i \sqrt{-\beta} b_2\right), & \beta<0,\end{cases} \\
h_2\left(\alpha, b_1, b_2\right) & = \begin{cases}\theta\left(b_2-b_1\right) I_0\left(\sqrt{\alpha} b_1\right) K_0\left(\sqrt{\alpha} b_2\right)+\left(b_1 \leftrightarrow b_2\right), & \alpha>0, \\
\theta\left(b_2-b_1\right) I_0\left(\sqrt{-\alpha} b_1\right) K_0\left(i \sqrt{-\alpha} b_2\right)+\left(b_1 \leftrightarrow b_2\right), & \alpha<0.\end{cases}
\end{aligned}
\end{equation}

The jet function $S_t(x)$ resums the threshold logarithm $\ln^2x$ appearing the hard kernels to all orders and is parameterized as
\be{}
S_t(x) = \frac{2^{1+2a}\Gamma(3/2+a)}{\sqrt{\pi}\Gamma(1+a)}[x(1-x)]^a,
\en{}
with the parameter $a = 0.4$.

The Sudakov factors used in the text are defined by
\begin{equation}
\begin{aligned}
S_{a b}(t)= & s\left(\frac{M_{B_c}}{\sqrt{2}} r_c, b_1\right)+s\left(\frac{M_{B_c}}{\sqrt{2}} x_3, b_3\right)+\frac{5}{3} \int_{1 / b_1}^t \frac{d \mu}{\mu} \gamma_q(\mu)+2 \int_{1 / b_3}^t \frac{d \mu}{\mu} \gamma_q(\mu), \\
S_{c d}(t)= & s\left(\frac{M_{B_c}}{\sqrt{2}} r_c, b_1\right)+s\left(\frac{M_{B_c}}{\sqrt{2}} z, b\right)+s\left(\frac{M_{B_c}}{\sqrt{2}}\left(1-z\right), b\right)+s\left(\frac{M_{B_c}}{\sqrt{2}} x_3, b_1\right) \\
& +\frac{11}{3} \int_{1 / b_1}^t \frac{d \mu}{\mu} \gamma_q(\mu)+2 \int_{1 / b}^t \frac{d \mu}{\mu} \gamma_q(\mu), \\
S_{ef}(t)= & s\left(\frac{M_{B_c}}{\sqrt{2}} r_c, b_1\right)+s\left(\frac{M_{B_c}}{\sqrt{2}} z, b_3\right)+s\left(\frac{M_{B_c}}{\sqrt{2}}\left(1-z\right), b_3\right)+s\left(\frac{M_B}{\sqrt{2}} x_3, b_3\right), \\
& +\frac{5}{3} \int_{1 / b_1}^t \frac{d \mu}{\mu} \gamma_q(\mu)+4 \int_{1 / b_3}^t \frac{d \mu}{\mu} \gamma_q(\mu),\\
S_{gh}(t)= & s\left(\frac{M_{B_c}}{\sqrt{2}} z, b\right)+s\left(\frac{M_{B_c}}{\sqrt{2}}\left(1-z\right), b\right)+s\left(\frac{M_{B_c}}{\sqrt{2}} x_3, b_3\right) \\
& +2 \int_{1 / b}^t \frac{d \mu}{\mu} \gamma_q(\mu)+2 \int_{1 / b_3}^t \frac{d \mu}{\mu} \gamma_q(\mu),
\end{aligned}
\end{equation}
where $K_0$ and $I_0$ are modified Bessel functions with $K_0(i x)=\frac{\pi}{2}\left(-N_0(x)+i J_0(x)\right)$ and $J_0$ is the Bessel function. The hard scale $t_i$ is chosen as the maximum of the virtuality of the internal momentum transition in the hard amplitudes

\begin{equation}
\begin{aligned}
t_a & =\max \left\{M_{B_c} \sqrt{\left|\alpha_a\right|}, M_{B_c} \sqrt{\left|\beta_a\right|}, 1 / b_3, 1 / b_1\right\}, && t_b=\max \left\{M_{B_c} \sqrt{\left|\alpha_b\right|}, M_{B_c} \sqrt{\left|\beta_b\right|}, 1 / b_1, 1 / b_3\right\}, \\
t_c & =\max \left\{M_{B_c} \sqrt{\left|\alpha_c\right|}, M_{B_c} \sqrt{\left|\beta_c\right|}, 1 / b_1, 1 / b\right\}, && t_d=\max \left\{M_{B_c} \sqrt{\left|\alpha_d\right|}, M_{B_c} \sqrt{\left|\beta_d\right|}, 1 / b_1, 1 / b\right\}, \\
t_e & =\max \left\{M_{B_c} \sqrt{\left|\alpha_e\right|}, M_{B_c} \sqrt{\left|\beta_e\right|}, 1 / b_3, 1 / b\right\}, && t_f=\max \left\{M_{B_c} \sqrt{\left|\alpha_f\right|}, M_{B_c} \sqrt{\left|\beta_f\right|}, 1 / b, 1 / b_3\right\},\\
t_g & =\max \left\{M_{B_c} \sqrt{\left|\alpha_g\right|}, M_{B_c} \sqrt{\left|\beta_g\right|}, 1 / b, 1 / b_1\right\}, && t_h=\max \left\{M_{B_c} \sqrt{\left|\alpha_h\right|}, M_{B_c} \sqrt{\left|\beta_h\right|}, 1 / b, 1 / b_1\right\},
\end{aligned}
\end{equation}

where
\begin{equation}
\begin{aligned}
& \alpha_a=r_b^2-(\zeta+x_3(1-\zeta))(1-r^2),                                   &&            \beta_a=((1-x_3)(1-\zeta)-r_c)(r_c-r^2), \\
& \alpha_b=(1-\zeta-r_c)(r^2-r_c),                                              &&            \beta_b=\beta_a \\
& \alpha_c=\beta_a,                                                             &&            \beta_c=-((1-\zeta)(1-x_3)+\zeta-r_c)((1-r^2)(1-z)+r^2-r_c), \\
& \alpha_d=\beta_a,                                                             &&            \beta_d=-((1-\zeta)(1-x_3)-r_c)((1-r^2)z+r^2-r_c) ; \\
& \alpha_e=-zx_3(1-\zeta)(1-r^2),                                                             &&            \beta_e=r_b^2-(1-x_3(1-\zeta)-r_c)(1-x_2(1-r^2)-r_c), \\
& \alpha_f=\alpha_e,                                                             &&            \beta_f=r_c^2-(r_c-x_3(1-\zeta))(r_c-z(1-r^2)). \\
& \alpha_g=-(\zeta+x_3(1-\zeta))(1-r^2),                                         &&      \beta_g=\alpha_e, \\
& \alpha_h=r_c^2-(1-\zeta)(z(1-r^2)+r^2),                                  &&            \beta_h=\alpha_e. \\
&
\end{aligned}
\end{equation}

\end{document}